\documentclass[journal, 12pt, onecolumn, draftclsnofoot] {IEEEtran}

\newcommand*{\OnlineSM}{}

\usepackage{pdfpages}
\usepackage{orcidlink}
\usepackage{amsmath,amsfonts}
\usepackage{array}
\usepackage[caption=false,font=normalsize,labelfont=sf,textfont=sf]{subfig}
\usepackage{textcomp}
\usepackage{stfloats}
\usepackage{url}
\usepackage{verbatim}
\usepackage{subfig}
\usepackage{float}
\usepackage[]{graphicx}
\usepackage{bmpsize}

\ifCLASSOPTIONcompsoc
  \usepackage[nocompress]{cite}
\else 
  \usepackage{cite}
\fi

\usepackage[libertine]{newtxmath}
\usepackage[nofillcomment]{algorithm2e}
\usepackage{algpseudocode}
\algrenewcommand\textproc{}
\usepackage{paralist}
\usepackage{upgreek}
\usepackage{float}
\usepackage{svg}
\usepackage{multirow}
\usepackage{array}
\usepackage{diagbox}
\usepackage{mathtools}
\usepackage{tikz,xcolor}
\usepackage{hyperref}
\hypersetup {colorlinks=false, linkcolor=blue, filecolor=blue, citecolor = black, urlcolor=cyan} 

\renewcommand{\thetable}{\Roman{table}}
\graphicspath{{Figures/png} {Figures/eps}}

\begin{document}


\title{An Efficient Hash-based Data Structure for Dynamic Vision Sensors and its Application to Low-energy Low-memory Noise Filtering}
\ifCLASSOPTIONcompsoc

\author{Pradeep~Kumar~Gopalakrishnan~\orcidlink{0000-0001-6597-6141},~\IEEEmembership{Senior Member,~IEEE,}~Chip-Hong Chang~\orcidlink{0000-0002-8897-6176},~\IEEEmembership{Fellow,~IEEE,}~and~Arindam Basu~\orcidlink{0000-0003-1035-8770},~\IEEEmembership{Senior Member,~IEEE.}

\IEEEcompsocitemizethanks{\IEEEcompsocthanksitem Pradeep Kumar Gopalakrishnan is with CNRS@CREATE LTD, 1 Create Way, \#08-01 CREATE Tower, Singapore 138602.

\IEEEcompsocthanksitem Chip-Hong Chang is with CNRS@CREATE LTD and Nanyang Technological University, Singapore. 

\IEEEcompsocthanksitem Arindam Basu is with City University of Hong Kong, Hong Kong.} 


}

\IEEEaftertitletext{\vspace{-2\baselineskip}} 
   \IEEEtitleabstractindextext{
\begin{abstract}

Events generated by the Dynamic Vision Sensor (DVS) are generally stored and processed in two-dimensional data structures whose memory complexity and energy-per-event scale proportionately with increasing sensor dimensions. In this paper, we propose a new two-dimensional data structure ($BF_2$) that takes advantage of the sparsity of events and enables compact storage of data using hash functions. It overcomes the saturation issue in the Bloom Filter (BF) and the memory reset issue in other hash-based arrays by using a second dimension to clear $1$ out of $D$ rows at regular intervals. A hardware-friendly, low-power, and low-memory-footprint noise filter for DVS is demonstrated using $BF_2$. For the tested datasets, the performance of the filter matches those of state-of-the-art filters like the BAF/STCF while consuming less than $10\%$ and $15\%$ of their memory and energy-per-event, respectively, for a correlation time constant $\uptau=5$ ms.  The memory and energy advantages of the proposed filter increase with increasing sensor sizes. The proposed filter compares favourably with other hardware-friendly, event-based filters in hardware complexity, memory requirement and energy-per-event---as demonstrated through its implementation on an FPGA. The parameters of the data structure can be adjusted for trade-offs between performance and memory consumption, based on application requirements. 

\end{abstract}
}
\maketitle
\else

\author{Pradeep~Kumar~Gopalakrishnan~\orcidlink{0000-0001-6597-6141},~\IEEEmembership{Senior Member,~IEEE,}~Chip-Hong Chang~\orcidlink{0000-0002-8897-6176},~\IEEEmembership{Fellow,~IEEE,}~and~Arindam Basu~\orcidlink{0000-0003-1035-8770},~\IEEEmembership{Senior Member,~IEEE.}

\thanks{Pradeep Kumar Gopalakrishnan is with CNRS@CREATE LTD, 1 Create Way, \#08-01 CREATE Tower, Singapore 138602.}
\thanks{Chip-Hong Chang is with CNRS@CREATE LTD and Nanyang Technological University, Singapore.}
\thanks{Arindam Basu is with City University of Hong Kong, Hong Kong.}


} 

\IEEEaftertitletext{\vspace{-2\baselineskip}} 

\maketitle
   
   \maketitle
\fi
\section{Introduction} \label{Introduction} The Dynamic Vision Sensor (DVS)---also known as Event-based Vision Sensor or Neuromorphic Vision Sensor---belongs to a special class of bio-inspired image sensors that output \textit{events} in response to changes in the intensity of light falling on them \cite{Delbruck2008}. Pixels in a DVS camera generate events asynchronously and independently when the temporal contrast changes (increases or decreases) by a set threshold. Ideally, if the light falling on DVS pixels does not change, they will not produce any output, thus providing background removal for stationary cameras at the pixel plane. This data reduction feature makes it attractive for applications of machine vision systems such as intelligent transportation and smart farming enabled by the Internet of Things (IoT)\cite{Mohan2022}. Moreover, since their pixels respond to changes in illumination asynchronously, the DVS can have very short latencies and several kHz of \textit{effective frame rate}. This feature, combined with a higher dynamic range ($\approx$$120$ dB), makes it a good candidate for applications such as high-speed motion estimation \cite{Benosman2012,Delbruck2013,Benosman2014}, robotic tracking \cite{Delbruck2013,Mueggler2014, Glover2017}, autonomous driving \cite{Maqueda2018}, drone navigation\cite{Gallego2019,Vidal2017}, high-speed and high dynamic range (HDR) imaging \cite{Rebecq2019, Rebecq2021}. Further, the low power consumption and high frame rates of the DVS makes it attractive for human-machine interaction \cite{Amir2017, Tapiador2020} as well.


Events generated from a DVS are typically encoded and transmitted using the Address Event Representation (AER) protocol; \cite{Mahowald1994} each event is represented by a quadruplet $e(x,y,t,p)$ consisting of:

\begin{compactitem}
\item $x$ - Column number of the pixel generating the event
\item $y$ - Row number of the pixel generating the event
\item $t$ - An $n_T$-bit timestamp (typically $32$-bit) of the event
\item $p$ - Polarity of the event ($1$ for \textit{ON} events---indicates a positive change in illumination, $0$ for \textit{OFF} events---indicates a negative change in illumination)
\end{compactitem}

\begin{figure} [htbp]
\centering
\includegraphics [width=\columnwidth, bb= -100 50 1000 220]{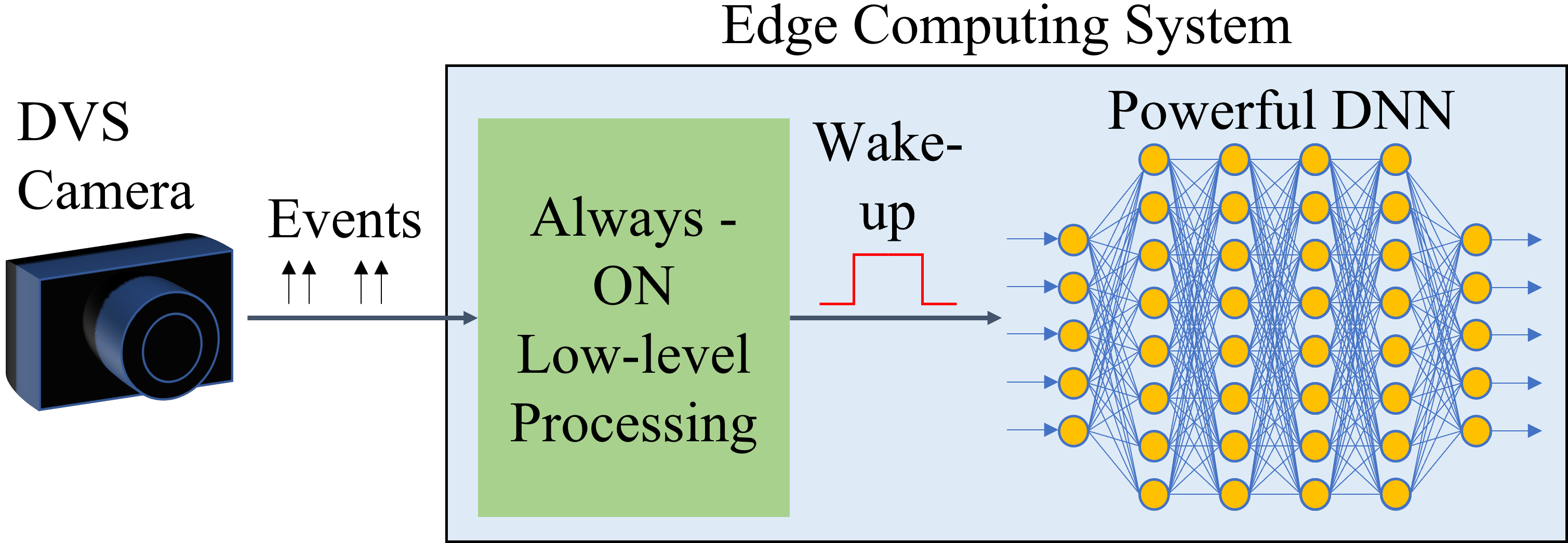}
\caption{Edge computing system with "Always-On" circuits.}
\label{fig:edge}
\end{figure}

\IEEEpubidadjcol To retain the advantages arising from the sparsity of events emitted by a DVS, it is necessary to store and process them efficiently. Early work with the DVS used serial processing on CPUs\cite{Delbruck2013,Everding2018} with a \mbox{$1$-D} event queue like the data structure. However, such \mbox{$1$-D} structures while being low on memory make the search for spatial neighbours (required for many algorithms such as noise filtering and corner detection) difficult. Some works\cite{Liu2018AdaptiveTB,Kogler2009,Mohan2022} have created frames or \mbox{$2$-D} histograms out of the events for further processing; however, the fine temporal resolution is lost as a result. By far, the most popular data structure used to process events is a time surface (TS)\cite{Delbruck2008,Lagorce2017,Manderscheid2019} or variants thereof\cite{Glover2021}. All these data structures require storing an $n$-bit timestamp (typically $n=32$) for all the pixel locations, signifying a large memory overhead. While the sparsity of events implies rare reads and writes to the memory, the energy required for each memory access is high due to the large size of the memory\cite{Horowitz2014}. Since energy dissipation is a prime consideration in the design of systems for IoT, it is  necessary to reduce the memory required to store the events. This is especially true in the case of ``Always-On" systems\cite{Bose2022} where there is the possibility of a more powerful processor being triggered to ``wake up" for further detailed processing. Fig. \ref{fig:edge} shows an example of this scheme.\IEEEpubidadjcol

%

Other data structures such as the octree, or time aggregation-based accumulation of spike events have recently been used to compress DVS data for transmission \cite{Dong2019,Khan2021}; however, they need to be decompressed before being processed further. In this paper, we propose a data structure with high space-efficiency and variable temporal resolution to facilitate the application of the DVS to energy and memory constrained edge computing systems. Specifically, the main contributions of this paper are: 
\begin {itemize}
\item A hardware-friendly, hashing-based data structure with constant memory complexity to store events from the DVS as well as search for the presence of past events is proposed.
\item A Background Activity (BA) removal filter is implemented using this data structure; the filter's requirements of memory, computing resources, and energy are low enough to enable its integration into the peripheral circuits of the DVS.

\item A design space exploration is performed to demonstrate the versatility of the data structure, and theoretical estimates of errors due to the memory being finite are provided.

\item It is demonstrated that the proposed filter can produce results comparable to or better than those of other hardware-oriented filters, while consuming lesser energy per event. 
\item  Results from a hardware implementation of the filter on an FPGA are presented.
\end {itemize}

The rest of the paper is organized as follows: Section \ref{Preliminaries} reviews related work in the domain. 
Section \ref{Data_Structure} describes the proposed data structure, with its application to BA filtering described in Section \ref{Proposed Filter}; the performance of the proposed filter for different memory configurations are theoretically analyzed, and the parameters for achieving optimum performance are predicted. Section \ref{Results} discusses the performance evaluation methodology, datasets, and the evaluation metrics for the filter, as well as the results of comparison of the proposed filter with similar filters in terms of performance and resource requirements. Section \ref{Hardware} describes the hardware implementation details of the filter.

\section{Preliminaries and Review of Related Works} \label{Preliminaries}
\subsection{Data structures for event storage and processing} \label{sec:prior_data_structures} 


Individual events do not convey much information about the scene when viewed in isolation; it is necessary to aggregate and store a set of related events in a suitable data structure before meaningful information can be extracted from them. A few such data structures have already been mentioned in Section \ref{Introduction}. In this section, we review some data structures that are commonly used to store and analyze event data  \cite{Gallego2019, Gehrig2019}.


\subsubsection{Event Frame / Event Image / \texorpdfstring{$2$-D}{TEXT} Histogram} \label{event_frame}
This data structure is a \mbox{$2$-D} image or ``frame" based on the count of events output by the sensor. Each pixel of the sensor corresponds to a specific location in the event frame. It is possible to have a combined count for both event polarities or two separate counts---one for each polarity---per pixel. The accumulation of counts can be for a fixed period of time or for a fixed number of events.

Event Based Binary Image (EBBI) \cite{Acharya2019, Mohan2022} is a variation of the event frame. It is possibly the simplest \mbox{$2$-D} data structure that can be created from an event stream. Each pixel in the EBBI is a single bit that indicates the presence or absence of an event in the corresponding sensor pixel during the period of accumulation of event counts.

Event frames have several advantages \cite{Gallego2019}:

\begin{compactitem}

\item {They have the ability to provide an adaptive sampling rate proportional to the activity in the scene.}

\item {They contain information on the presence and absence of events.}

\item {They convert the event stream to a more familiar \mbox{$2$-D} image that includes the edges of the scene, and thus provide an intuitive interpretation of the information in the scene.}

\item {In some applications, event frames may be directly processed by conventional (frame-based) image processing algorithms (e.g. nearest neighbor filtering).}
\end{compactitem}

Due to these advantages, several researchers have used this data structure in their works \cite{Rebecq2017, Kogler2009, Liu2018ABMOF, Maqueda2018, Gehrig2020}. However, event frames merge the  temporal information from individual events into a single frame. Thus, the temporal resolution of the information contained in the event stream is lost, nullifying one of the main benefits of asynchronous address-event driven data processing that differentiates it from its frame-based counterparts.

\subsubsection{Time surface (TS)} \label{Time Surface}

Like the event frame, a TS is also a \mbox{$2$-D} array that has the same dimensions (number of pixels in the $x$ and $y$ dimensions) as the sensor. It stores the most recent timestamp of an event generated from a sensor pixel, at a particular $(x,y)$ location \cite{Delbruck2008, Lagorce2017}. A TS stores only one timestamp per pixel, which is updated asynchronously. Modified versions of the basic TS that are more tolerant to variations in motion speed are described in \cite{Manderscheid2019} and \cite{Alzugaray2018}. Filtering of events in a space-time window to make the TS more immune to noise is proposed in \cite{Sironi2018}.

\subsubsection{Threshold Ordinal Surface (TOS)}

TOS is a \mbox{$2$-D} data structure introduced in \cite{Glover2021}, where it is used to detect corners in an event data stream based on an implementation of the Harris corner detection algorithm \cite{Harris1988}. The TOS is an ``image" array of dimensions $R \times C$, where $R$ and $C$ are the sensor dimensions (the number of rows and number of columns of the pixel array respectively). TOS contains values between $0$ and $255$, on which the Harris corner detection algorithm is executed. Every incoming event modifies a part of the TOS using an update algorithm, based on two parameters $k_{TOS}$ and $T_{TOS}$ whose values are determined based on the application scenario.

\subsubsection {\texorpdfstring{$3$-D}{TEXT} point set}
In this approach, a spatio-temporal neighborhood is considered, and events enclosed within this space $(x_k,y_k, t_k) \in \mathbb{R}^3$ are treated as points in a \mbox{$3$-D} space. Here, the time dimension is converted to a spatial dimension (the $z$ dimension). This yields a sparse \mbox{$3$-D} ``event cloud" data structure on which point-based methods can be applied to derive properties such as the visual flow \cite{Benosman2014}.

\subsubsection {Point sets on image plane}
In this method, events are treated as a set of \mbox{$2$-D} points on the image plane that evolve over time. This representation is created by shape tracking methods based on the Iterative Closest Point (ICP) \cite{Kueng2016} or mean shift \cite{Litzenberger2006, Delbruck2013}  algorithm.

\subsubsection{Voxel grid}
A voxel grid is a \mbox{$3$-D} representation of events in a spatio-temporal neighborhood. Each voxel represents an event within a specified time interval. Voxel grids preserve timing information better than  \mbox{$2$-D} structures such as event frames \cite{Zhu2018}.

\subsubsection{Bloom Filter}

The Bloom Filter (BF) is a probabilistic data structure that can be used to efficiently test the membership of an element in a set \cite{Bloom1970}. It uses $k$ hash functions to encode the input element to $k$ integer values. These $k$ values are then used as indices pointing to a bit-array of $m$ cells, all of which are set to zero at the start of the filtering process. When any index points to a cell of the memory array, that cell is set to one. Elements are presented to the BF one by one. If all the $k$ cells indexed by the $k$ hash functions return ones, the element is considered to be a member of the set; this means that it was presented to the BF earlier. The BF has been used in \cite{Guo2021} to implement a filter for event-based data. It is however noted that the BF faces a problem of saturation when a large percentage of the bits happen to be set to ones, leading to the possibility of a query returning a false positive. The equation for the False Positive Rate of a BF is given in Section \ref{FPR_theory}. 



Table \ref{tab:ds_memory} shows the comparison of memory complexities of the different data structures.
It can be seen from the table that most of the data structures used for event storage and processing have a memory complexity of $O(R \times C)$. This makes it
difficult to implement them in hardware with low complexity and energy efficiency, as explained further in Section \ref{BA_noise_filtering}. Although the BF has a constant memory complexity, it cannot be directly used with continuous event data streams due to its susceptibility to saturation, which results in a high number of false positives. 
In this work, we propose a data structure that can efficiently store and retrieve information from event data streams, and has the constant-order memory characteristics of the BF.

\renewcommand{\arraystretch}{1.4} 

\begin{table}[htbp]
\centering
\caption{Comparison of memory complexity of data structures}
\setlength{\tabcolsep}{4.5pt}
\renewcommand{\arraystretch}{1.25}
\label{tab:ds_memory}
\begin{tabular}{|l|c|l|l|}
\hline
\textbf{Data structure} & \multicolumn{1}{l|}{\textbf{\begin{tabular}[c]{@{}l@{}}Order of\\ Memory   \\ Complexity\end{tabular}}} & \textbf{\begin{tabular}[c]{@{}l@{}}Memory\\ requirement\\ (bits)\end{tabular}} & \textbf{Remarks} \\ \hline
\begin{tabular}[c]{@{}l@{}}Event frame /\\ Event image /\\ \mbox{$2$-D} Histogram\end{tabular} & \multirow{7}{*}{$O(R \times C)$} & \begin{tabular}[c]{@{}l@{}}$R \times C \times n_C$ or\\ $2 \times R \times C \times n_C$\end{tabular} & \begin{tabular}[c]{@{}l@{}}$n_C$ is the \\ bit-width of \\ the event count.\end{tabular} \\ \cline{1-1} \cline{3-4} 
EBBI &  & $R \times C$ &  \\ \cline{1-1} \cline{3-4} 
Time Surface (TS) &  & $R \times C \times n_T$ & \begin{tabular}[c]{@{}l@{}}$n_T$ is the \\ bit-width of \\ the timestamp.\end{tabular} \\ \cline{1-1} \cline{3-4} 
\begin{tabular}[c]{@{}l@{}}Threshold \\ Ordinal \\ Surface (TOS)\end{tabular} &  & $R \times C \times 8$ & \begin{tabular}[c]{@{}l@{}}TOS is specially \\ designed for \\ corner detection.\end{tabular} \\ \cline{1-1} \cline{3-4} 
3-D point set &  & $R \times C \times N_e$ & \begin{tabular}[c]{@{}l@{}}$N_e$ is the average \\ number of events  \\ enclosed in the \\ 3-D volume.\end{tabular} \\ \cline{1-1} \cline{3-4} 
\begin{tabular}[c]{@{}l@{}}Point sets\\ on image plane\end{tabular} &  & $R \times C$ &  \\ \cline{1-1} \cline{3-4} 
Voxel Grid &  & $R \times C \times B$ & \begin{tabular}[c]{@{}l@{}}$B$ is the number \\ of temporal bins.\end{tabular} \\ \hline
Bloom Filter & $O(1)$ & $W$ & \begin{tabular}[c]{@{}l@{}}$W$ is the number \\ of bits in the\\ Bloom filter.\end{tabular} \\ \hline
\end{tabular}
\end{table}


\subsection{Background Activity (BA) Noise}\label{BA Noise}
Though the DVS has several advantages, a major disadvantage is the generation of BA noise events even without any change in illumination. These events are a result of temporal noise, and junction-leakage currents in the circuitry of the sensor pixels \cite{Lichtsteiner2008}, \cite{Liu2015}, with a higher level of BA noise expected in low-light situations \cite{Lichtsteiner2008},\cite{Hu2021}. BA noise events can adversely impact the quality of data output from the DVS, result in wastage of communication bandwidth, and cause unnecessary energy consumption in downstream stages of processing. It is important to filter out these events at the earliest possible stage in the event processing chain---ideally, at the sensor or at the always-ON block during edge processing. In this paper, we propose a filter based on a new data structure that requires minimal memory and hardware resources, suitable for implementation in Always-On edge processing (Fig. \ref{fig:edge}). 


\subsection{BA Noise Filtering} \label{BA_noise_filtering}
\begin{figure} [htbp]
\centering 
\includegraphics[width=\columnwidth, bb = -50 20 600 300]{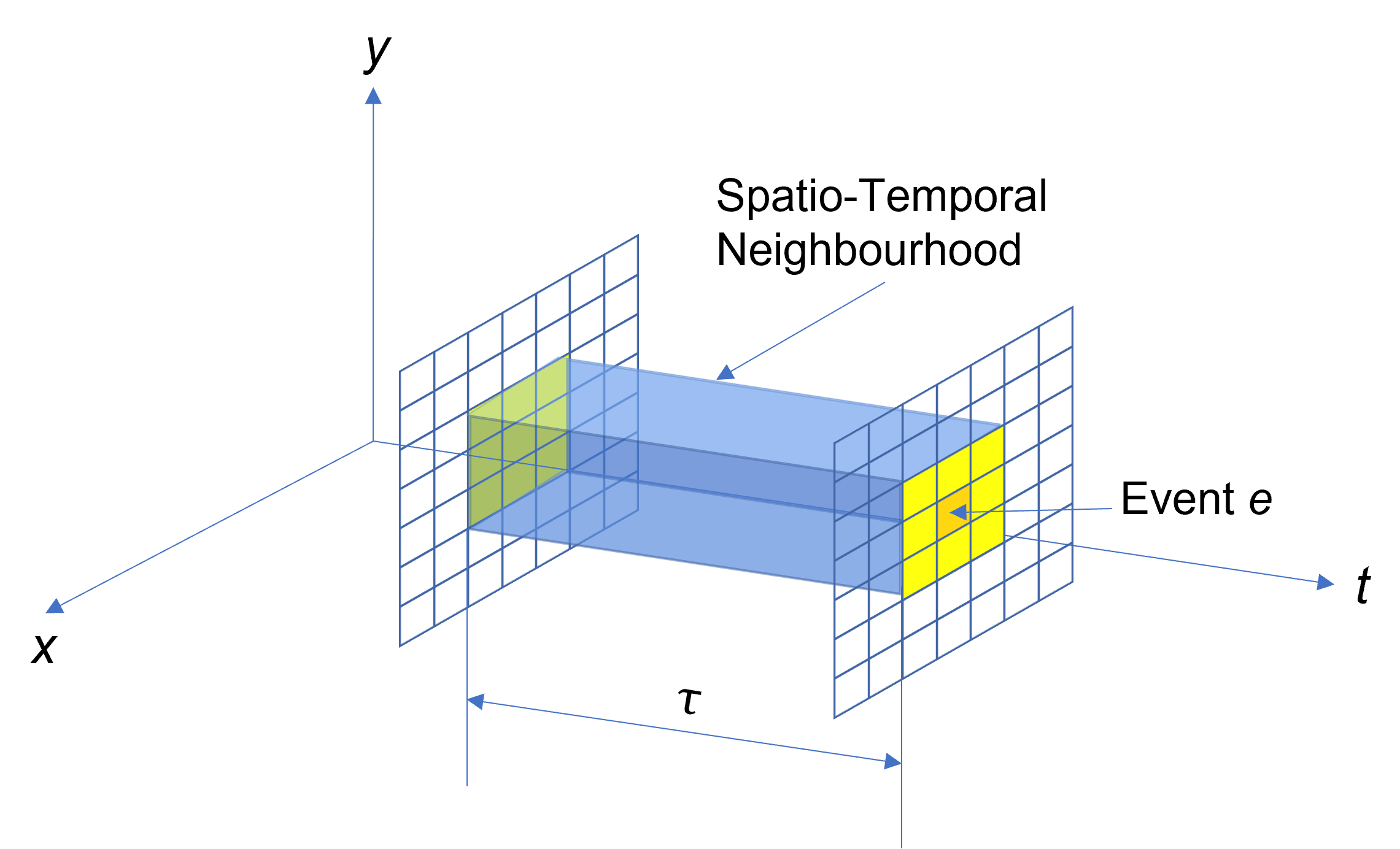}
\setlength{\belowcaptionskip}{-10pt}
 \caption{The STCF principle: At least $k$ events are required in a spatio-temporal neighbourhood for declaring an event as signal.}
\label{fig:stcf}
\end{figure}

BA noise events usually occur in isolation, and are generated at random pixel locations and time instants. In contrast, events resulting from real activity usually occur in groups of several adjacent pixels, close together in time. Hence, filtering of BA events can be achieved by blocking out events that lack spatio-temporal correlation to other events. The process of separating BA noise events from real events based on this property is known as Spatio-Temporal Correlation Filtering (STCF). The STCF principle, illustrated in Fig. \ref{fig:stcf}, forms the basis of many event-based filters. The grids represent pixel locations in the sensor. The \textit{spatio-temporal neighborhood} of the current event $e$ is indicated in blue. This is a space-time volume that encloses events occurring adjacent to $e$ (within a neighbourhood of $\pm n$ pixels in the vertical or horizontal direction), within a constant \textit{correlation time} duration of $\uptau$ from its timestamp. Any event that occurs within this space-time volume is called a supporting event of $e$. When an event $e$ arrives, it is classified as signal if at least $s$ supporting events are found, otherwise it is classified as noise. Note that the neighborhood does not include the location of $e$ itself to prevent noise events from \textit{hot pixels}---pixels that fire repeatedly at the same location---from escaping the filter.

\subsubsection{Background Activity Filter} \label{BAF}
Probably the most popular implementation of STCF---which we refer to as the Background Activity Filter (BAF)---is presented in \cite{Delbruck2008}. In this filter, a TS data structure is used, and the parameters of the STCF are set to $n=s=1$. The BAF is effective in removing BA noise events even from complex scenes.  A related filter proposed in \cite{Guo2022} (we refer to this filter as Guo-STCF) is a generalization of the BAF discussed in Section \ref{BAF}. In this filter, $n=1$ as before but $k>1$ to achieve better filtering. 

\begin{figure} [htbp]
\centering 
\includegraphics[width=\columnwidth, bb = -100 50 1200 600]{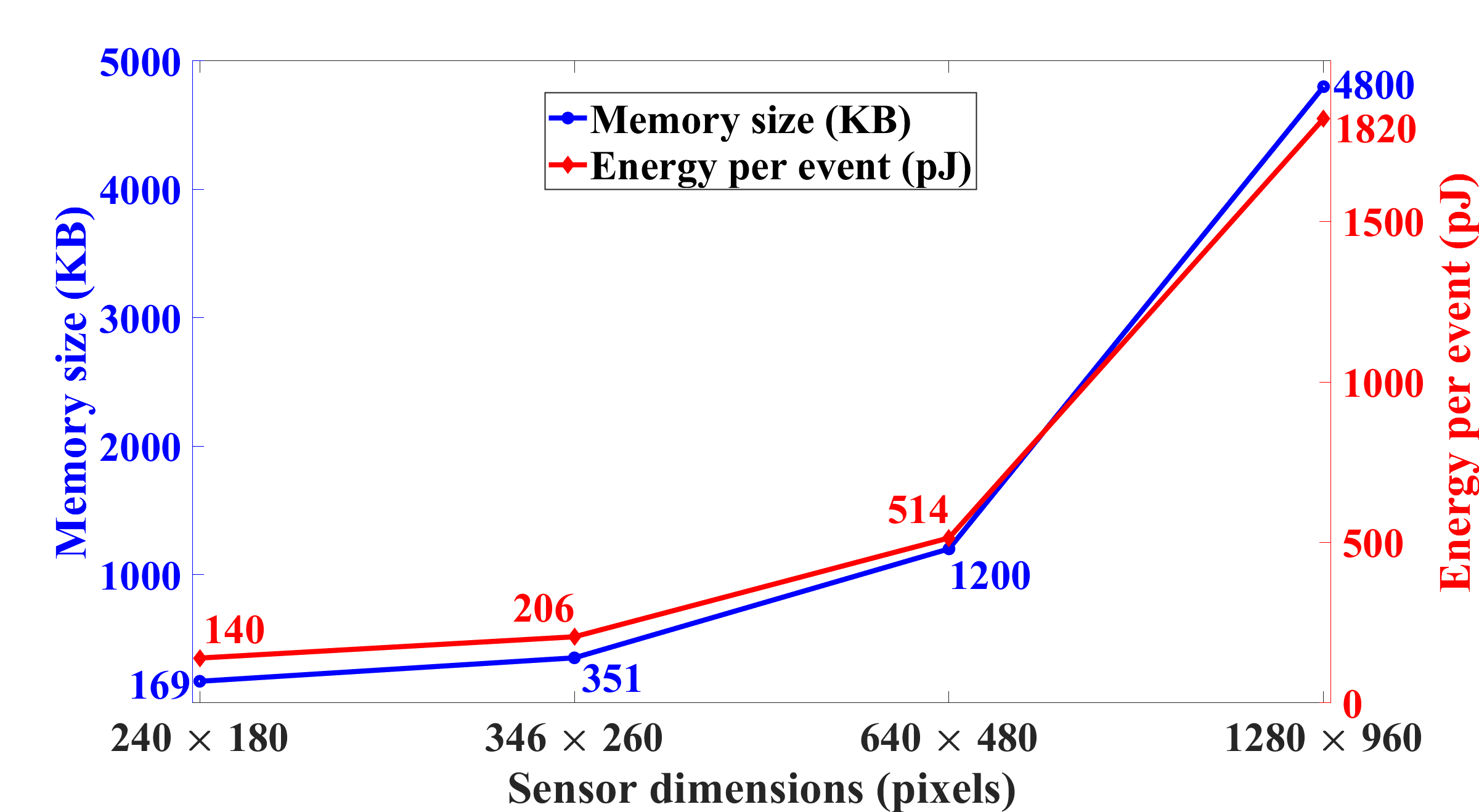}
 \caption{Memory size and energy per event vs Sensor dimensions: Due to use of the TS data structure, memory size and energy per event increase proportionally with the sensor size for the BAF or Guo-STCF.}
\label{fig:baf_pj_mem}
\end{figure}

However, the memory requirement of these filters is very high ($R \times C \times n_T$ bits)  
due to the TS array used in them, with the memory requirement increasing with increasing sensor resolution, as shown in Fig. \ref{fig:baf_pj_mem}. As the memory size increases, the energy cost of accessing a \textit{single bit} of the memory array increases \cite{Horowitz2014}, leading to a rapid escalation in energy cost per event as well. This is confirmed by  our estimates (based on \cite{Horowitz2014}) of energy cost per event for different sensor dimensions, as shown in Fig. \ref{fig:baf_pj_mem}. Note that a higher resolution of the sensor is accompanied by an increase in event rate as well, leading to an even higher rate of power dissipation. This highlights the problems associated with using data structures with memory complexity $O(R \times C$) in low-power, always-ON modules.

\subsubsection{O(N) Filter (ONF)} \label{ONF}
A filter with memory requirement linearly correlated to the sensor size is proposed in \cite{Khodamoradi2021}. In this filter, two $n_T=32$-bit memory cells are allocated corresponding to each row and each column of the sensor array, resulting in a memory requirement of $(R + C)\times 64$ bits. 
The information on the $x$, $y$, $p$, and $\emph{ts}$ parameters related to the latest event in the row or column are stored in these cells. 
To classify an incoming event as signal or noise, the memory cells corresponding to its neighboring row and column are read to check for the presence of supporting events in its spatio-temporal neighborhood. Although this filter has a very low memory footprint, the filter's performance was found to be lower than that of other filters in our simulations, especially in filtering complex datasets (Section \ref{perf_comp}). Moreover, the data structure provides no flexibility to improve the performance by adding more resources.


\subsubsection{HashHeat Filter} \label{hashheat}
HashHeat is a hashing-based BA filter proposed in \cite{Guo2021}. This filter has a memory complexity independent of the sensor size. It uses a combination of the BF (described in Section \ref{sec:prior_data_structures}) and a set of $k$ Locality Sensitive Hash (LSH) functions to index the BF array. The LSH function equation is given by \cite{Guo2021}: 
\begin {equation} \label{lsh_function}
h_{i}= {((a_{i} \cdot x+b_{i} \cdot y+c_{i} \cdot \emph{ts}+d_{i})}/w)) \bmod m
\end {equation}
where $x$, $y$, and $\emph{ts}$ are the coordinates of the latest event and
its timestamp respectively, $a_i$, $b_i$, $c_i$, and $d_i$ are random numbers, $m$ is the number of cells in the BF memory array and $w$ is a parameter called segment length. 

The memory array can be single-bit or multi-bit, depending on the complexity of the dataset used. The memory array is reset once after every $N$ events. The coordinates and timestamp of an input event are used by the $k$ LSH functions to generate indices used to access the memory array. The outputs of the memory array are compared with a threshold $\emph{Thr}$ to classify the event as signal or noise. Since the LSH function needs to be evaluated for each incoming event, its complexity is an important factor in the filter's energy efficiency. In Supplementary Material
\ifdefined \OnlineSM
(SM)---\href{https://sites.google.com/view/bf2-event-based-filter/supplementary-material/section-a}{SM Section A},
\else
(SM)---SM Section A,
\fi
we compare the area and power requirements of the LSH function implemented in a $40$ nm CMOS process with those of the hash function used in the proposed filter.
Although HashHeat is frugal in its memory use and efficient in implementing the STCF search, it has the disadvantage that the relationship between the various filter parameters and the resulting filter performance is not well-defined; how to choose these parameters to achieve a particular correlation time constant $\uptau$ is therefore unclear.

\subsubsection{Event Density Based Filter}
In \cite{Feng2020}, a method for filtering the BA events based on event density is proposed. The filtering proposed in this work involves two stages, with the first one filtering random noise and the second one, hot pixels. The method involves  accumulating the number of events in a space-time window around each incoming event to form an ``event density matrix" (same as the event frame mentioned in Section \ref{event_frame}), and then finding its L2 norm. This can be expensive to implement in hardware; no hardware implementation is presented in the paper.

\subsubsection{Other Hardware-friendly Filters}
Four noise filtering algorithms are proposed in \cite{Guo2022}. Guo-STCF is one of them while the other two are lightweight algorithms named Fixed Window Filtering (FWF) and Double Window Filtering (DWF). These filters are only effective for low-complexity and sparse activity filtering situations with stationary cameras. Hence we do not consider them in our filter comparisons. The last one is a filter based on neural networks called Multi-layer Perceptron de-noising Filter (MLPF). It exploits structural cues in the local spatio-temporal window to improve filtering performance; however, no hardware implementation of MLPF is presented. The basic data structure that forms the input to the neural network being a TS (Section \ref{Time Surface}), has a memory complexity of $O(R \times C)$. 
 
A Neural Network-Based Nearest Neighbor (NeuNN) filtering algorithm based on neuromorphic integrate-and-fire neurons is proposed in \cite{Padala2018}. However, this work is based on a specialised neuromorphic processor IBM TrueNorth, which is not yet commercially available.

BA filtering is employed in \cite{Acharya2019} as a part of an  object tracker for event-based cameras. The method uses the EBBI data structure, and performs median filtering on them. Although this can reduce computations, the creation of frames from events results in the loss of many advantages that the event-based paradigm mentioned in Section \ref{Introduction} provides. Also, it has been proven to work only for simple situations involving stationary cameras.

Most of the BA filters mentioned previously are either too resource-hungry to implement in/near the sensor of an event camera or are only effective in filtering low-complexity, sparse-event streams from static event cameras. Our work aims to develop a low-complexity filter that is effective in applications that use mobile cameras producing event streams of high complexity. In the following sections, we present our new memory-efficient data structure as an extension of the BF, followed by a BA filter based on it that is able to effectively filter complex event streams from moving cameras combined with heavy noise under low-light conditions.

\section{Proposed Hash-based Data Structure: \texorpdfstring{$BF_2$}{BF2}} \label{Data_Structure}

As mentioned in Section \ref{sec:prior_data_structures}, most of the data structures currently used for event storage and processing have a memory complexity of $O(R \times C)$, making them non-optimal for low-power, low-cost hardware implementations. Also, though the BF has the advantage of presenting the least memory complexity, it cannot be used directly for processing event data streams due to the possibility of saturation. The related data structure used in HashHeat\cite{Guo2021} solves the saturation problem with regular reset operations, but suffers from data loss after reset. This prompted us to propose a new hash-based data structure for efficiently storing and retrieving event information within a limited temporal neighborhood similar to a two-dimensional BF but without data loss during reset. The data structure was designed to be hardware friendly, with a constant-order memory requirement and low energy consumption.

The proposed data structure, denoted by $BF_2$, comprises a set of $K$ \mbox{$2$-D} arrays, $BF_2^1$ to $BF_2^K$, each with width $W$ and depth $D$. Insertion of an event $e(x,y,t,p)$---denoted by $RowPtr=InsertEvent(BF_2,RowPtr,x,y,t)$---into $BF_2$ is described by Algorithm \ref{alg:insert_event}.

\RestyleAlgo{ruled}
\SetKwComment{Comment}{/* }{ */}
\begin{algorithm}[htb]
\scriptsize 
\caption{Event insertion} \label{alg:insert_event}
\tcc {Inserts the input event into the BF$_2$ array}
\tcc {Returns the latest value of RowPtr}
\begin{algorithmic}
\Function{{InsertEvent}}{$x,y,t,RowPtr,\uptau_{row},W,D,K$} \\
\tcc {Identify the row for storing the input event}
$R \leftarrow (\lfloor\frac{t}{\uptau_{row} }\rfloor \mod{D}) + 1$; \\
\tcc{Check the condition to clear the row}
\If {$(R \neq RowPtr)$} {
\tcc{Clear bits in the newly-active row prior to inserting new events}
$BF_2^i(R,j) \leftarrow 0, j = 1,2,\ldots, W; i=1,2,\ldots, K$
}
\tcc{Prepare the indices for setting bits}
$Idx_i \leftarrow hash_i(x,y),i=1,2,\ldots,K$\\
\tcc {Set bits in the active row}
$BF_2^i(R,Idx_i) \leftarrow 1, i=1,2,\ldots,K$ \\
\tcc{Return the updated value of RowPtr}
\Return{$R$}
\EndFunction
\end{algorithmic}
\end{algorithm}

In Algorithm \ref{alg:insert_event}, $h_i()$ denote hash functions within the range $1$--$W$. $RowPtr$ denotes a row pointer pointing to the currently ``active" row, and is incremented once after every time period $\uptau_{row}$ (the time allotted to each row for storing events); when $RowPtr$ reaches a value of $D$, the row pointer rolls back to $1$. Before writing to a new row once $RowPtr$ is incremented, the row is cleared to make room for new entries. The rows of the array enable storage of events presented to it as well as their retrieval, working like a BF. The search operation, denoted by $[D_{out}(1),D_{out}(2),..D_{out}(D)]=SearchEvent(BF_2,x,y)$, is described by Algorithm \ref{alg:search_event}.

\RestyleAlgo{ruled}
\SetKwComment{Comment}{/* }{ */}
\begin{algorithm}[htb]
\scriptsize 
\caption{Event search} \label{alg:search_event}
\tcc {Searches the BF$_2$ array for the presence of existing events with input coordinates $(x,y)$}
\tcc{Returns the value of BF$_2$ row outputs}
\begin{algorithmic}
\Function{{SearchEvent}}{$x,y,D,K$} \\
\tcc {Prepare the indices for search}
$BitPtr_i \leftarrow Hash_i(x,y),i=1,2,\ldots,K$ \\
\tcc{Store the array outputs temporarily}
$Bit_{out}(i,j) \leftarrow BF_2^i(j,BitPtr_i), j=1,2,\ldots,D; i=1,2,\ldots,K$ \\
\tcc{Combine outputs from all hashes}
$D_{out}(j) \leftarrow AND(Bit_{out}(1,j), Bit_{out}(2,j),\ldots,Bit_{out}(K,j))$ \\
\tcc {Return BF2 row outputs}
\Return{$D_{out}(j), j=1,2,\ldots,D$}
\EndFunction
\end{algorithmic}
\end{algorithm}

In Algorithm \ref{alg:search_event}, $hash_i()$ denote the hash functions as before. In this algorithm, the search outputs from the $BF_2$ arrays are indicated by $Bit_{out}(i,j)$, where $1\leq i\leq K$ corresponds to the different hash functions and $1\leq j\leq D$ is the row number. The logical AND applied to the outputs from a given row ``$j$" across multiple hash tables is used to generate a signal $D_{out}(j)$ that indicates the presence or absence of the event at the queried location in that row ($1$ for presence and $0$ for absence); $D$ such outputs (i.e., one output for each row) are returned by the search operation. This AND operation helps reduce the return of false positives during search in $BF_2$ arrays. 

Essentially, this data structure creates a sliding temporal window of size $D \times \uptau_{row}$ in which the latest events from an event stream are stored. Within this window, there are several timing ``bins" of size (time interval) $\uptau_{row}$, with each event stored in one of the bins based on its temporal position in the event stream. Thus, the information regarding the timing of the events is not entirely discarded, but quantized and stored. The resolution of this quantization can be adjusted based on the application. Each row of the $BF_2$ array corresponds to one temporal bin. Once stored, the data structure  enables a user to efficiently search for the presence of past events with a specified $(x, y)$ location and locate it within a specific bin (row) in the temporal window. The rate of false positives generated by the search process can be controlled by varying the amount of memory allotted to the arrays, while the temporal resolution of the  bins can be controlled by varying $\uptau_{row}$.

\begin{figure}[htb]
\centering
\captionsetup[subfigure]{font=scriptsize,labelfont=scriptsize}
\subfloat[]{\includegraphics[width=\paperwidth, bb=-200 0 1000 300] {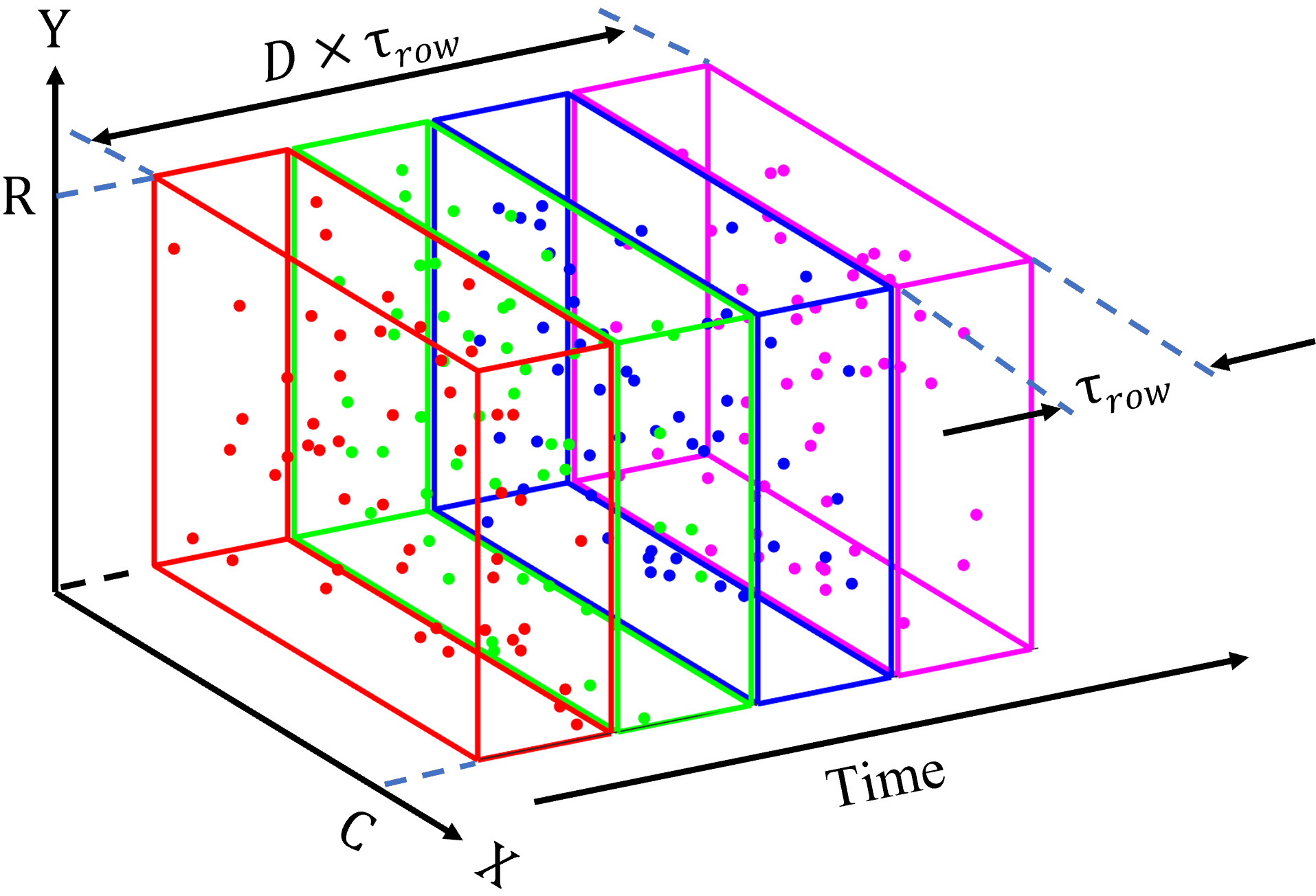}}

\subfloat[]{\includegraphics[width=\paperwidth, bb=-100 0 1000 300] {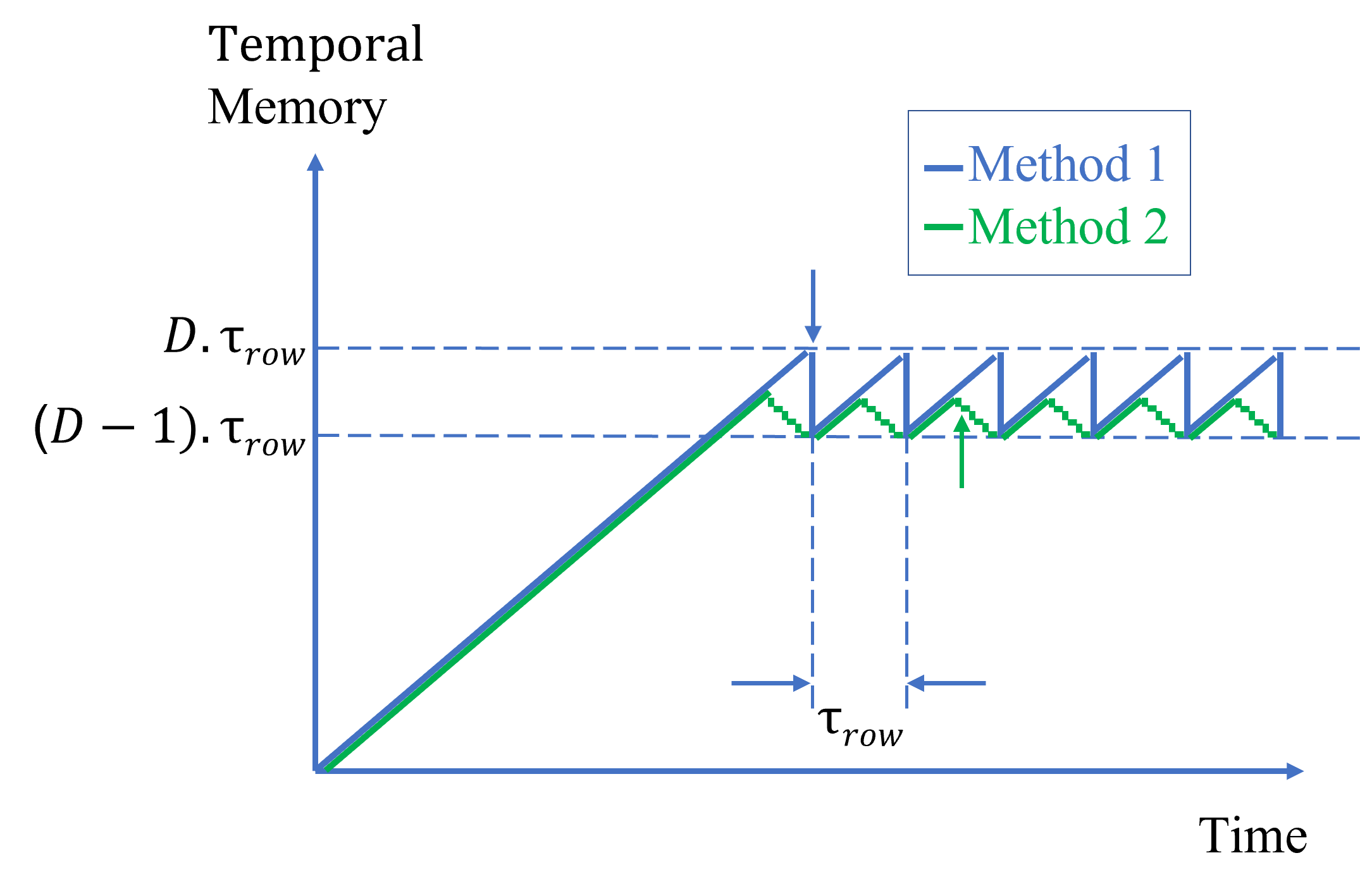}}
\caption{(a) Temporal memory and timing bins of the proposed $BF_2$ array corresponding to each row and spanning across $\tau_{row}$. (b) Two methods of clearing rows to prevent saturation: Method 1 - delete row (N) after writing into row (N-1); Method 2 - delete row (N) while writing into row (N-1).}

\label{fig:timing_bins}
\end{figure}

Fig. \ref{fig:timing_bins}(a) illustrates the concept of the temporal window and timing bins. This figure shows four timing bins corresponding to four rows of the $BF_2$ array. Events are represented by the colored dots and the timing bins by the colored boxes. The events belonging to a timing bins are given the same color as the bin.


In order to prevent the $BF_2$ array from saturating and resulting in a large false positive rate for the searches, it is necessary to clear the oldest events from the array in a continuous manner. These events are found in the least recently active row of the array, which is the row next in sequence to the currently active one. This process, shown in Fig. \ref{fig:timing_bins}(b) can be performed in two ways. In the first method ( Method 1 in the figure),  writes to the current row are completed after which the next row is cleared. In the second method (Method 2 in the figure), the next row in sequence is cleared \textit{while} the current row is being written into. The second method is applicable for hardware implementations that cannot clear a complete row of memory all at once. This method is further detailed in Section \ref{mem_reset}.


The total memory requirement of the proposed data structure is $K \times W \times D$ bits. The data structure can be used in any application where there is a requirement to search for the presence of past events (using their $x$, $y$ coordinates) in an event stream, within a limited time period following the latest event. Examples of such applications are corner detection and filtering.

Next, we illustrate how the proposed data structure can be used to implement a BA filter for DVS cameras.
\section {DVS noise filtering using \texorpdfstring{$BF_2$}{BF2} arrays} \label{Proposed Filter}

\subsection{STCF realization using \texorpdfstring{$BF_2$}{BF2}}

Section \ref{BA_noise_filtering} shows that the classification of an incoming event as signal or noise based on the STCF principle consists of a search process involving events falling within a time interval of $\uptau$ before the event. For every incoming event to be classified as signal or noise, a search needs to be performed to check for the presence of past supporting events in terms of $x,y$ coordinates, in the neighbourhood of the current event. If the event has at least $k$ supporting events within the time duration $\uptau$ just prior to the current event, it is classified as signal, else it is classified as noise.

This classification can be readily accomplished with the proposed data structure using the following steps:

\begin{enumerate} 
    \item 
    Set the timing window size to $\uptau$ by setting time per row $\uptau_{row}=\uptau / D $ so that $\uptau  = \uptau_{row} \times D$.
    \item 
    Search for the locations of the eight immediate neighbours of the current event in $BF_2$. A controller can be used to generate the $x,y$ coordinates of these neighbours to be input to the data structure.
    \item 
    For each search, perform the logical OR on the outputs from all rows of the data structure. This is needed as the STCF principle looks for supporting events anywhere within the time duration $\uptau$. If the result of the OR operation is $1$, it signifies the presence of a supporting event, and the current event is classified as a signal.
\end{enumerate}

\begin{figure} [ht]
\centering 
\includegraphics[width=\paperwidth, bb=0 40 1200 500] {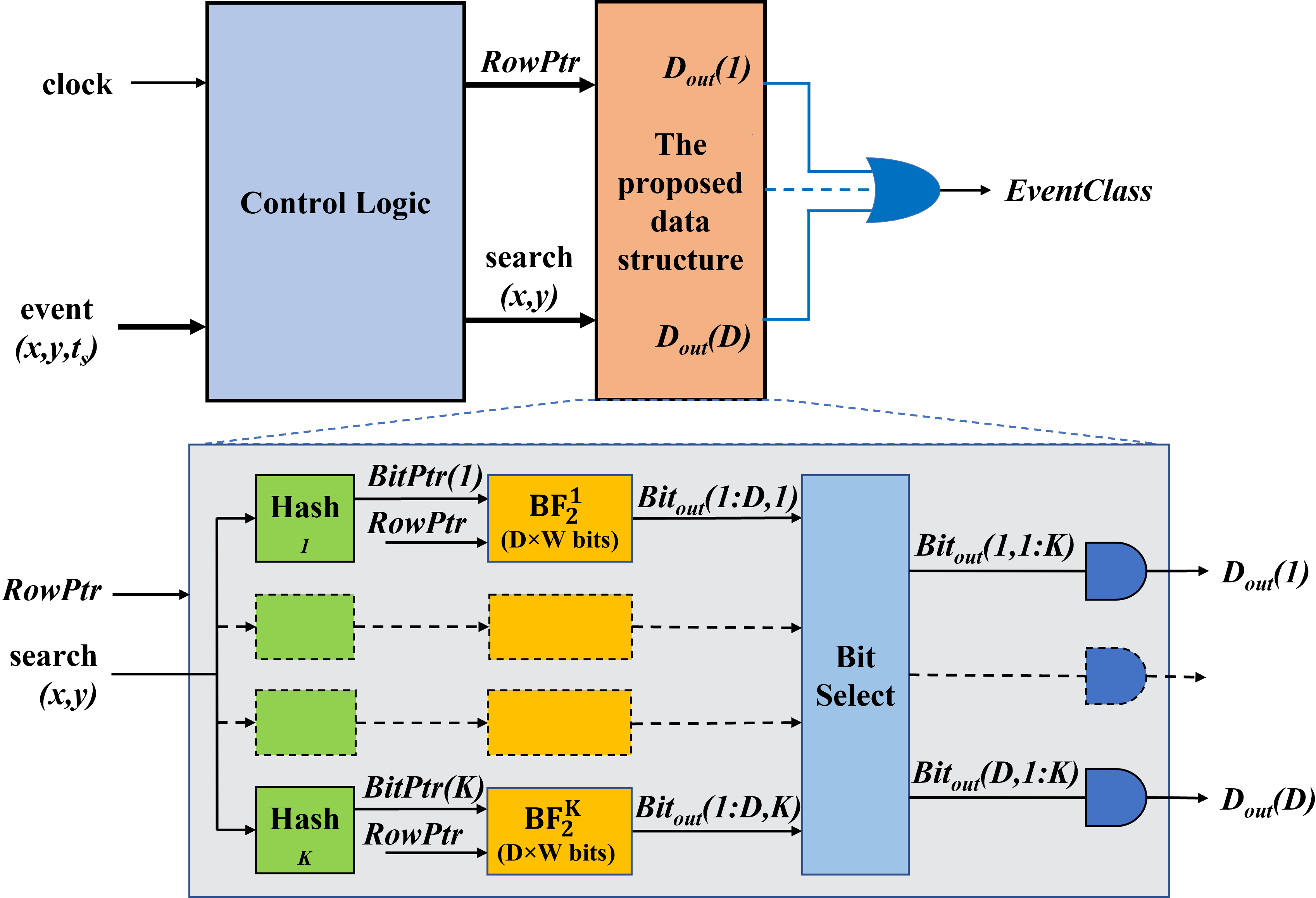} \caption{Structure of the proposed noise or background activity filter for DVS using $BF_2$ arrays.}
\label{fig:filter_struct}
\end{figure}

A block diagram representing a filter based on the proposed data structure, that uses the above method is shown in Fig. \ref{fig:filter_struct}. The corresponding filtering algorithm is shown in Algorithm \ref{alg:classification}.

\RestyleAlgo{ruled}
\SetKwComment{Comment}{/* }{ */}
\begin{algorithm}[htb]
\scriptsize 
\caption{Event classification} \label{alg:classification}
\KwIn {$x,y,t,D,K$}
\KwOut{$EventClass,RowPtr$}
$EventClass \leftarrow 0;$\\
\tcc {Search for supporting events in the neighborhood of the current event, in all rows of all BF$_2$ arrays}
\For{m = $-1$ \KwTo $1$} {
    \For{n = $-1$ \KwTo $1$} {
        $D_{out}(j) \leftarrow SearchEvent(x+m,y+n,D,K); j=1,2,\ldots,D$
    }
}
    
\tcc {Insert the current event into the BF$_2$ arrays for future searches, update  $RowPtr$ according to the timestamp of the latest event}
$RowPtr \leftarrow InsertEvent (x,y,t,RowPtr,\uptau_{row},W,D,K);$

\tcc{$EventClass$ is the logical OR of the data outputs of all rows of the BF$_{2}$ arrays. If the input event is classified as signal, EventClass $\leftarrow$ 1, else EventClass $\leftarrow$ 0}

$EventClass \leftarrow OR(D_{out}(1),D_{out}(2),\ldots,D_{out}(D))$ 
\end{algorithm}

In hardware implementations, the outer-most loop of this algorithm involving $K$ FIFOs can be run concurrently to save computation time. The process to update $RowPtr$ is run at every tick of the filter input clock. 



\subsection{Theoretical analysis of filter performance}\label{Theoretical Analysis of filter}

\subsubsection{Filter performance metrics} \label{FPR,TPR,F1}
An event-based filter can be considered a binary classifier since it classifies each incoming event as either signal or noise. The output of the filter can be compared with a pre-defined `Ground Truth' (GT) to determine the accuracy of the classification process.
False Positives (FP) are events classified by the filter as signal but are specified as noise in the GT while False Negatives (FN) are events classified by the filter as noise but specified in the GT as signal. 

The parameters False Positive Rate (FPR) and False Negative Rate (FNR) indicate the ratios of FP and FN to the total positive and negative samples, respectively. Also, $FP+TN = N_P$ is the total number of positive samples (signal) and $FN+TP = N_N$ is the total number of negative samples (noise), as defined by the GT.


Another parameter commonly used to measure filter performance is the F1 score that combines the effects of TP, FP and FN. It is given by $F1=\frac{2 \times Pr \times Re}{Pr + Re}$.
where $Pr$ is the Precision and $Re$ is the Recall as per their standard definitions.  

In the following sections, we derive equations that predict the performance of the filter. It is assumed that a recording sample from a scenario similar to that during deployment is available, along with the corresponding GT. These equations are useful for predicting the optimum filter memory configurations, as demonstrated in Section \ref{DSE}.

\subsubsection{Theoretical prediction of FPR} \label {FPR_theory}
As per the BF theory \cite{Bloom1970}, the FPR of the filter is given by:
\begin{equation} \label{eqn:Bloom_FPR}
FPR = \varepsilon \approx\ (1-e^{-k {\cdot} n/m})^{k}
\end{equation} 
where $n$ is the number of items stored in the array, $k$ is the number of hash functions, and $m$ is the number of bits in the bit array. Equation (\ref{eqn:Bloom_FPR}) holds good if $m \gg k$ \cite{Christensen2010, Bose2008}, as in our case where the number of hash functions $K$ is chosen to be much less than the array width $W$.

In our filter, the number of items (events) stored in a row of the $BF_2$ array denoted by $n_{row}$ can be approximated by the product of the time allotted to the row ($\uptau_{row}$) and the mean event rate ($\overline{e_{rate}}$), i.e. $n_{row}\approx \overline{e_{rate}}\times \uptau_{row}$. Assuming a constant $\overline{e_{rate}}$, the theoretical mean FPR of a \textit{single row} of a $BF_2$ array can be derived using (\ref{eqn:Bloom_FPR}):
\begin{equation}
\label{eqn:FPR_row}
FPR_{row}(n_{row}) =\left(1-e^{-\frac{n_{row}}{W}}\right)^{K} \approx \left(1-e^{-\frac{\overline{e_{rate}} {\cdot} \uptau_{row}}{W}}\right)^{K}
\end{equation}
Note that the exponent of $e$ in (\ref{eqn:FPR_row}) does not have the number of hashes ($K$) as a factor like in (\ref{eqn:Bloom_FPR}) since we use one memory row/bank per hash in our filter unlike the classical BF which uses one row for all the $k$ hashes.

The FPR of array $BF_2^i$ can be calculated as:

\begin{equation}
\label{eqn:FPR_dash_FIFO}
FPR_{BF_2}(n_{row}) = 1-(1-FPR_{row})^{D} 
\end{equation}
which indicates the probability of none of the $D$ rows having stored the event. Finally, the FPR for STCF filtering can be derived using the following equation, considering the probability of error in one of the $8$ neighbours:
\begin{equation}
\label{eqn:FPR_FIFO}
FPR_{STCF,BF_2}(n_{row}) = 1-(1-{FPR_{BF_2}})^8 \end{equation}

\begin{figure}[htb]
\centering
\captionsetup[subfigure]{font=scriptsize,labelfont=scriptsize}
\subfloat[]{\includegraphics[width=0.45 \paperwidth, bb=0 0 1700 800]{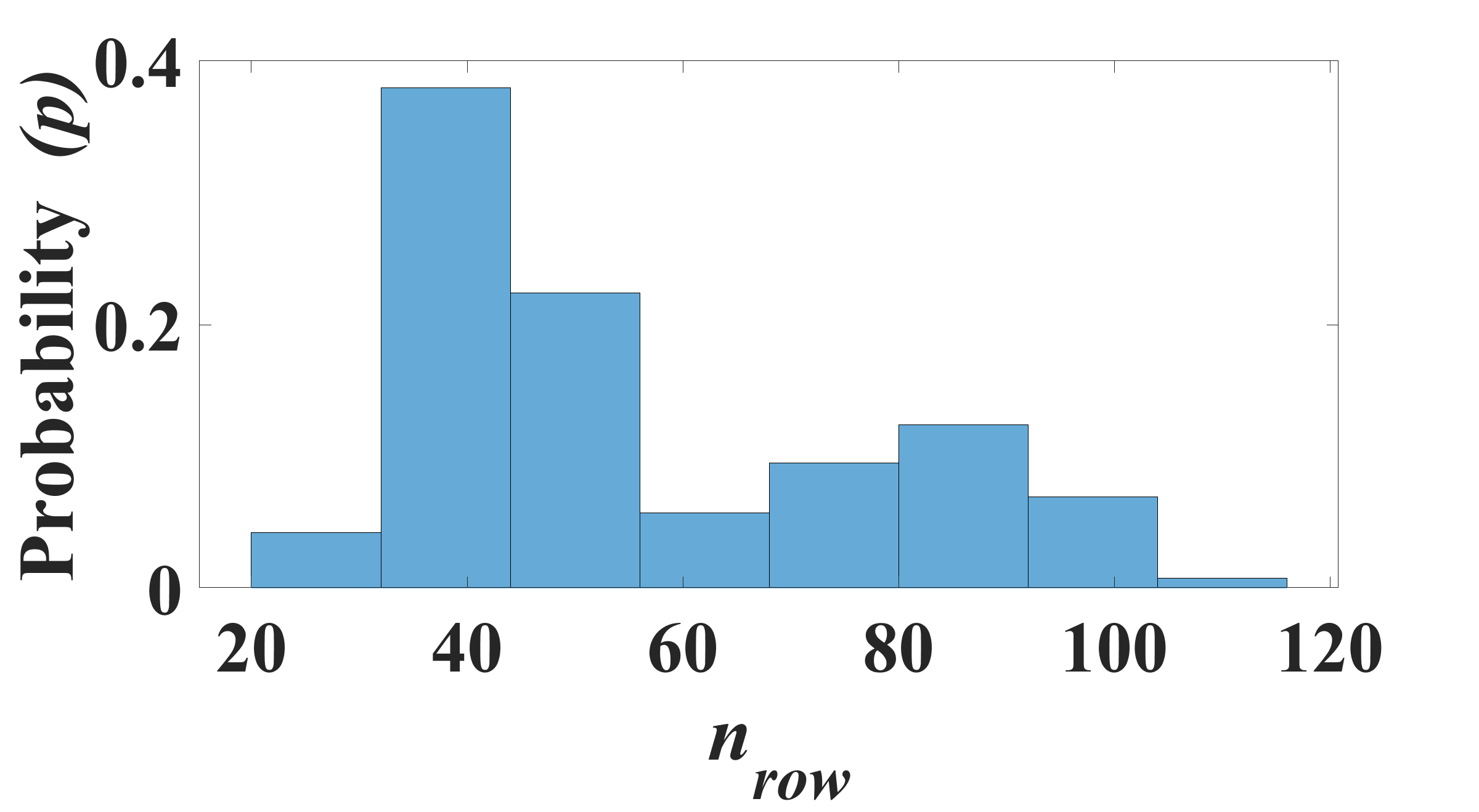}}
\subfloat[]{\includegraphics[width=0.45 \paperwidth, bb=0 0 4000 800]{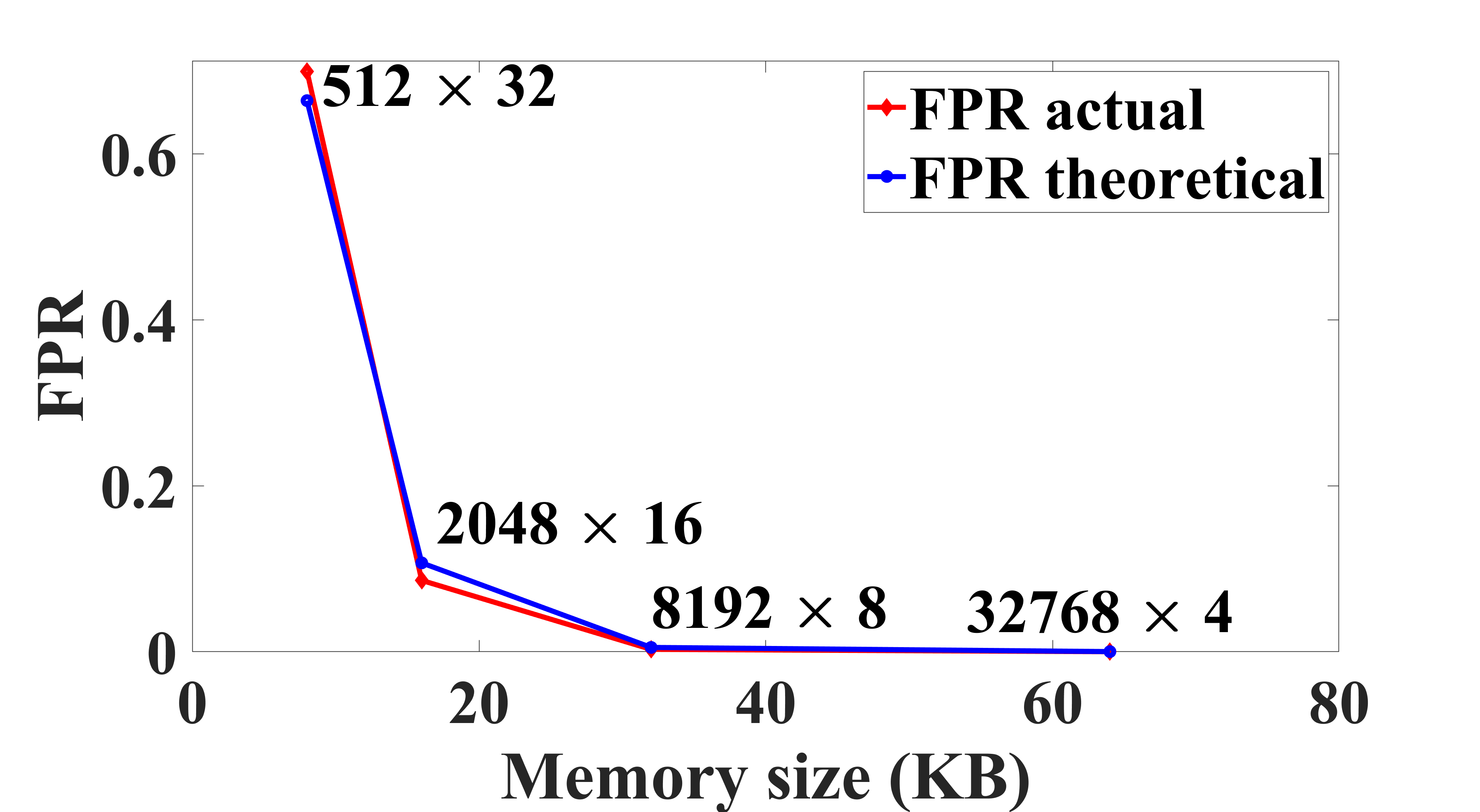}}

 \caption{(a) Example of histogram indicating number of events per $\tau_{row}$, and (b) comparison of FPR predicted by Equation (\ref{eq:weighted_FPR}) using these probabilities with actual FPR.}
\label{fig:fpr_hist}
\end{figure}

Equation (\ref{eqn:FPR_dash_FIFO}) accounts for the cumulative FPR of the $BF_2$ array, and depends on the number of rows $D$ while Equation (\ref{eqn:FPR_FIFO}) accounts for the worst case of eight searches in the neighbourhood of the event. The above equations were derived assuming constancy of $\overline{e_{rate}}$ and thereby constancy of $n_{row}$. However, in practice, the mean event rate from a DVS may vary significantly from time to time. Taking this into account, a more accurate estimate of the expected value of FPR can be calculated using weighted averages as follows:
\begin{equation}
\label{eq:weighted_FPR}
\langle FPR \rangle = \sum_{i} p(i) {\cdot} FPR_{BF_2}(i)
\end{equation}
where $p(i)$ denotes the probability that $n_{row}=i$. 
An example showing the probability distribution of $n_{row}$, for the `driving' dataset (described in Section \ref{Driving dataset}),  is seen in Fig. \ref{fig:fpr_hist}(a) .

The example in fig. \ref{fig:fpr_hist}(b)  shows the values of FPR predicted using this histogram. Here, the theoretically predicted FPR (using Equation \ref{eq:weighted_FPR}) is compared to the actual FPR determined from simulations, varying the total memory used by the filter from $8$ KB to $64$ KB (by varying $W$ and $D$), with $\uptau=5$ ms and the number of hash functions, $K=4$. The aspect ratios used for the different memory sizes ($W\times D$) are also indicated. It can be seen that there is a close correlation between the predicted and the actual FPR values.


\subsubsection{Theoretical prediction of FNR}
FNR for the classical BF is zero while in our filter, an FN is possible. This is because  a portion of the memory array is  periodically cleared to prevent saturation of the array. As mentioned in Section \ref{Proposed Filter}, events are transferred to the array row-by-row; when the time spent on one row exceeds $\tau_{row}$, the row next to it in sequence is cleared to prepare it to store future events. If the incoming event at any point of time has supporting events in the row being cleared, they may be misclassified as noise, resulting in FN.

The FNR is theoretically predicted based on the statistics of the number of neighbouring events within a support time of $\tau = D {\cdot} \tau_{row}$ before any event, support time being the time interval between an event and its support event in the past; the prediction is carried out for all the cases where the time difference $\Delta t$ between the current event and its neighbouring events is less than $\tau$ since these events would provide valid support for the STCF. From this distribution, the probability of support events falling in the last row being cleared can be estimated using the following equation:
\begin{equation}
\label{eq:FNR_th}
    \langle FNR \rangle =N_P {\cdot} \sum_{\mathclap{\Delta_t=(D-1) {\cdot} \tau_{row}}}^{D {\cdot} \tau_{row}} p(\Delta_t)
\end{equation}

\begin{figure}[htb]
\centering
\captionsetup[subfigure]{font=scriptsize,labelfont=scriptsize}
\subfloat[]{\includegraphics[width=0.45 \paperwidth, bb=0 10 4000 1300]{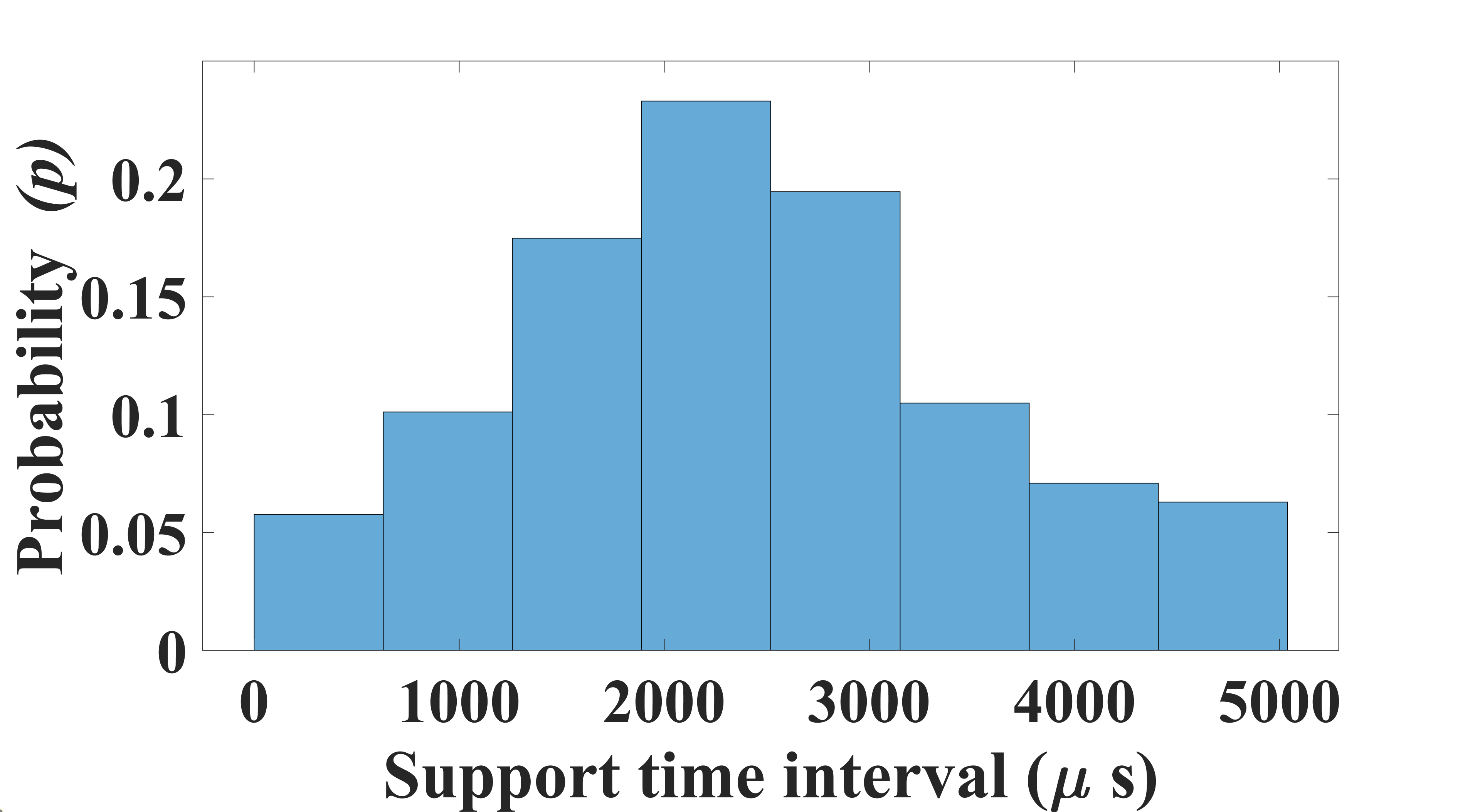}}
\subfloat[]{\includegraphics[width=0.45 \paperwidth, bb=0 10 4000 1300]{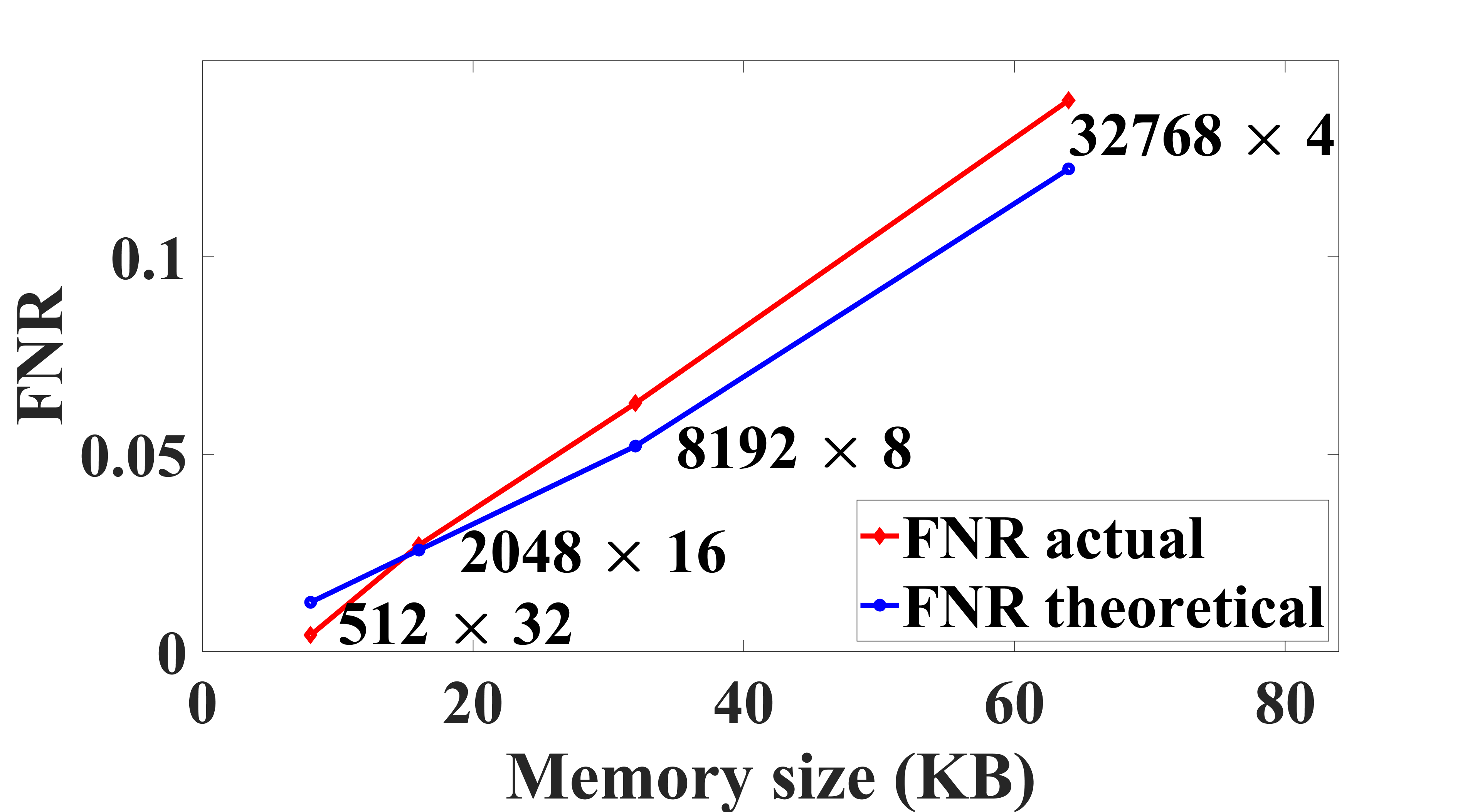}}
 \caption{(a) Estimation of FNR from a probability distribution of support time intervals, and (b) comparison of FNR predicted using these probabilities with actual FNR.}
\label{fig:fnr_hist}
\end{figure}

Fig. \ref{fig:fnr_hist}(a) is an example of such a probability distribution of $\Delta_t$ for the driving dataset. Fig. \ref{fig:fnr_hist}(b) compares the FNR predicted using Equation (\ref{eq:FNR_th}) with the actual FNR for various array configurations ($W\times D$) with different memory sizes; the comparison indicates a close match between the predicted and actual values.



\subsection{Design Space Exploration (DSE)} \label{DSE}

The proposed $BF_2$ data structure has several degrees of freedom such as the width $W$, depth $D$ and total memory $W\times D$. Utilizing the previously derived equations for $\langle FPR \rangle$ and $\langle FNR \rangle$, we may now perform a large-scale design space exploration (DSE) to explore the trade-offs between various parameters. Given a sample dataset and a target $\uptau$ value, we aim to determine the optimum memory size and aspect ratio ($W/D$) to be used, so as to achieve the best BA filtering effect.

The primary metric that we use to arrive at an optimum memory size and aspect ratio is the F1 score. Equations (\ref{eq:weighted_FPR}) and (\ref{eq:FNR_th}) from Section \ref{Theoretical Analysis of filter} can be combined to predict where F1 will peak in a sweep of different memory configurations for a given memory size. The value of $F1$ can be theoretically predicted by Equation (\ref{eqn:F1_th}) which uses the predicted values of $FPR$ and $FNR$ as shown below:
\begin{equation}
\label{eqn:F1_th}
F1 =\frac{2 {\cdot} N_P {\cdot} (1-\langle FNR \rangle)} {N_P {\cdot}  (2 - \langle FNR \rangle) + N_N {\cdot} \langle FPR \rangle}
\end{equation}
A detailed derivation of (\ref{eqn:F1_th}) is given in
\ifdefined \OnlineSM
\href{https://sites.google.com/view/bf2-event-based-filter/supplementary-material/section-b}{SM Section B}.
\else
SM Section B.
\fi
\begin{figure}[htb]
\centering
\captionsetup[subfigure]{font=scriptsize,labelfont=scriptsize}
\subfloat[]{\includegraphics[width=0.45\textwidth, bb=0 0 2800 1500]{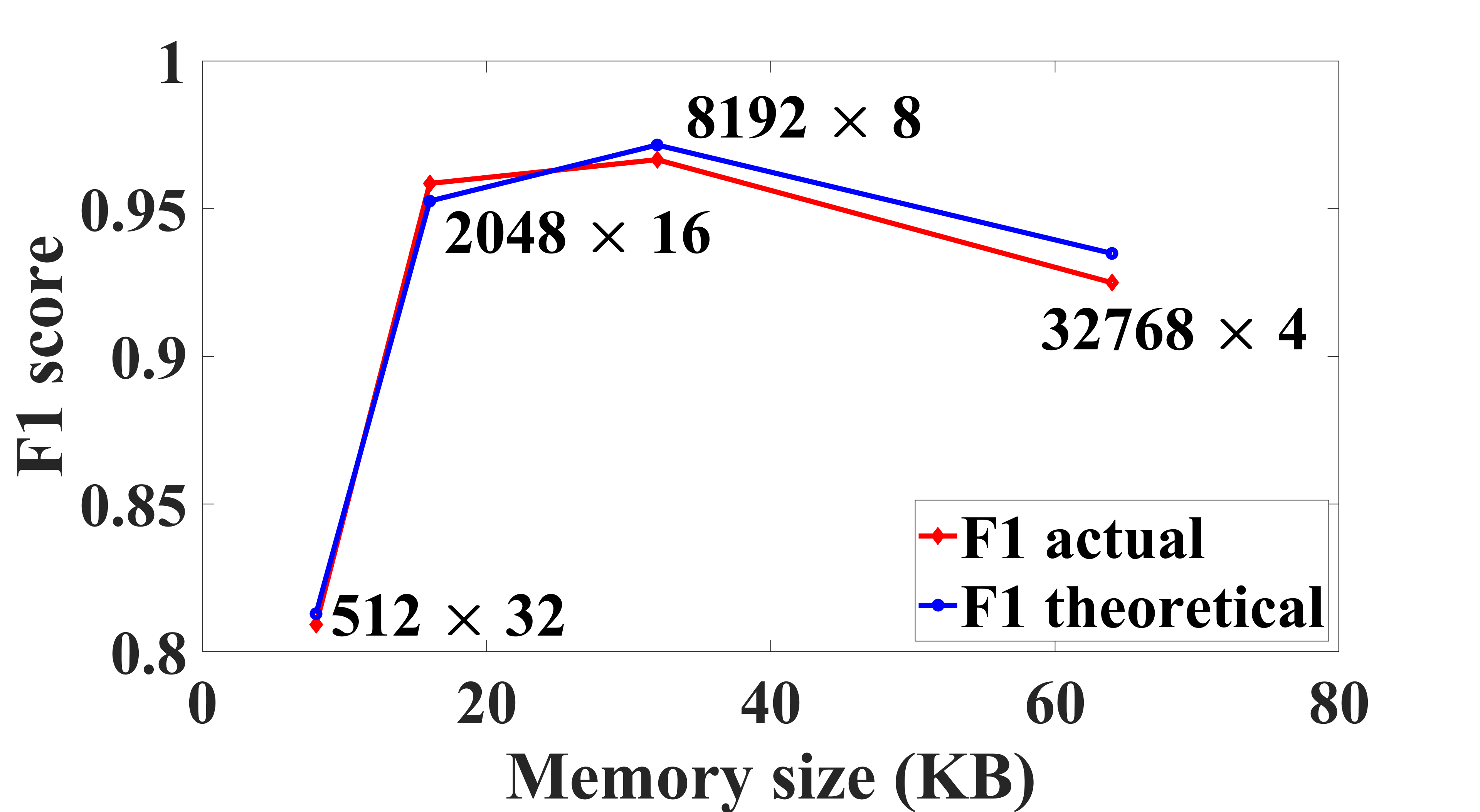}}
\subfloat[]{\includegraphics[width=0.45\textwidth, bb=0 0 2800 1500]{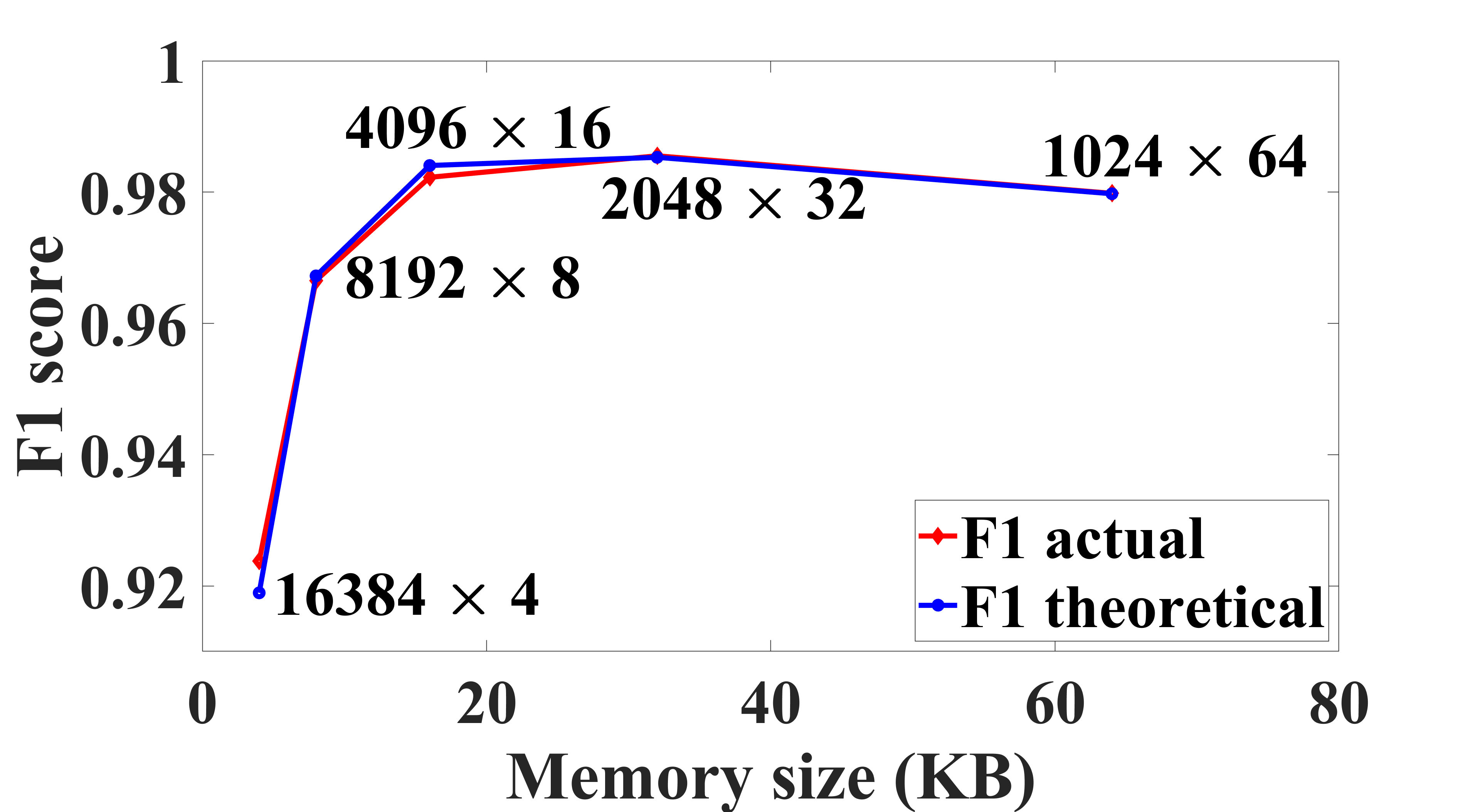}}
 \caption{(a) F1 curves (actual and theoretical) obtained by varying the total memory sizes, keeping the aspect ratio fixed. (b) F1 curves (actual and theoretical) for varying aspect ratios, at fixed total memory $W\times D$.}
\label{fig:f1_vs_f1_th}
\end{figure}


Fig. \ref{fig:f1_vs_f1_th}(a) shows a sweep of memory configurations for a memory size, and the resulting actual and theoretical F1 values based on Equation (\ref{eqn:F1_th}). It can be seen that it is possible to predict the optimum memory size (which is $8$ $\times$ $4$ = 32 KB for this example with the number of hash functions $K=4$) theoretically using the method outlined above. It may be noted that as seen in Fig. \ref{fig:f1_vs_f1_th}(a), we vary the aspect ratio of the memories as the memory size increases. This is because keeping the aspect ratio constant would lead to memories with large depth, resulting in a large FPR (this is evident from Equation (\ref{eqn:FPR_dash_FIFO})); to prevent this,  we increase the array width keeping the depth to a relatively low value when using larger memory sizes.

Once the optimum memory size is determined, it is possible to determine the optimum 'aspect ratio' of the memory array through a sweep of aspect ratios, keeping total memory size ($W \times D$) constant. An example of this is shown in Fig. \ref{fig:f1_vs_f1_th}(b) where the optimum aspect ratio is observed to be $1024$ $\times$ $64$.


\section{Results} \label{Results}

\subsection{Filter Evaluation Methodology} \label{Performance Evaluation Methodology}



The BA de-noising performance of the proposed filter was evaluated, and its comparison with other selected hardware-friendly filters was carried out using the MATLAB simulation environment. A MATLAB model of the proposed filter was created based on the architecture described in Section \ref{Proposed Filter}. Models of the other filters selected for comparison were created based on their description from \cite{Liu2015},\cite{Guo2021} and
\cite{Khodamoradi2021}. For the HashHeat filter\cite{Guo2021}, we used our best estimates of the filter parameters (mentioned in Section \ref{hashheat}).

We used two datasets to evaluate the proposed filter and to compare it with other filters. The performance evaluation flow is different for the two datasets used, mainly due to the difference in the way the GT is defined. The datasets and their corresponding performance evaluation flows are described next. 

\subsection{Datasets} \label{Datasets}
\subsubsection{Driving dataset}\label{Driving dataset}
The \textit{driving} dataset from DND21 \cite{Guo2022} was created by converting a simulated video (scenes captured from the dashboard camera of a car driving through a city) to event data using the Video to Events (v2e) tool \cite{Hu2021}. The resolution of the video is $346$ $\times$ $260$ pixels. It has a length of $\approx$ $6$ seconds and contains $\approx$ $3.9$ M events. This dataset represents a moving camera application of DVS that produces events from a substantial fraction of DVS pixels during a short time. 
Noise-free events from the driving dataset were mixed with recorded noise (also from DND21) to create the event data stream that was applied to the filter. Two different noise data files are available in the DND21 dataset; these are the \textit{leak noise} and \textit{shot noise} data files, recorded using a Davis346 camera under high and low intensity illumination conditions respectively. 
By mixing the contents of either of the recorded noise data files with the original noise-free data, the output of the camera operating under the corresponding lighting condition can be simulated.

In the performance evaluation setup that uses the driving dataset, the definition of GT considers all events from the original (noise-free) driving dataset as signals, and those from the recorded data as noise. Based on how an event is labelled by the filter and is defined in the GT, it can be classified as TP/TN/FP/FN. From this, the TPR and FPR values can be calculated, as described in Section \ref{Theoretical Analysis of filter}.

\subsubsection{Hotel-bar} \label {Hotel-bar dataset}
The \textit{hotel-bar} dataset from DND21 represents a typical surveillance operation using a stationary DVS camera, created by recording scenes from a hotel lobby using a DAVIS346 event camera and aggressively de-noising the camera output. The video has a resolution of $346$ $\times$ $260$ pixels and a length of approximately $12$ minutes. We used a shortened version of the video which has length of approximately $6.2$ seconds and containing $614$ K events. As with the \textit{driving} dataset, the noise-free video files were mixed with the \textit{leak noise} and \textit{shot noise} files before being applied to the filters, as described for the earlier dataset.

\subsubsection{NUS Traffic dataset} \label {NUS Traffic dataset} 
The NUS \textit{traffic} dataset discussed in \cite{Mohan2022} (available at \cite{traffic-dataset}) represents another stationary surveillance camera operation. It was acquired using a DAVIS240 event camera placed at a traffic junction, monitoring objects such as humans, motorcycles, cars, vans, trucks and buses moving on the road. The dataset includes GT boxes manually drawn around objects of interest in the scene, at specific time instances. Several short data files were created from the traffic data event stream by saving events in the temporal neighbourhood (within $\sim$$100$ ms) of the GT box time instants. Adopting the approach described in \cite{Padala2018}, the GT was defined by labelling events in these files as signal if they were projected to fall within the GT boxes, and as noise otherwise. These data files were input to the filter one by one, and their average FPR and TPR values were calculated.

\subsection{Filtering Performance} \label{perf_comp}

We used the methodology outlined in Section \ref{Performance Evaluation Methodology} to evaluate  the performance of the proposed filter and compare it with four other hardware-friendly filters discussed in Section \ref{Preliminaries}. The filters selected for comparison are the Guo-STCF, BAF, HashHeat and ONF. We use the Receiver Operating Characteristic (ROC)\cite{Fawcett2006} and Area under the ROC curve (AUC) as the primary parameters to evaluate the performance of the filters. The ROC curve is plotted by varying the correlation time constant $\uptau$ following the methodology in \cite{Guo2021}. 
The datasets described in Section \ref{Datasets} were used as input to the filters. For all filters except the HashHeat filter, the correlation time $\uptau$ was swept in the range of $100$ $\mu$s to $1$ s. 

For the HashHeat filter, the parameter $w$ in equation \eqref{lsh_function} was swept in the range $1$--$8192$. As for the the other parameters in the equation, $m$ = $4096$ ($8$-bit memory array), and $k$ = $4$ while parameters $a_{i},b_{i},c_{i}$ were assigned random values in the range $0$--$1$. These parameters were chosen from several trials so as to maximize filtering performance while minimizing memory usage.

In the case of our filter, the memory size that yields an F1 value close to (within $5$\%) those from the BAF was chosen corresponding to each value of $\uptau$. The memory allocated for the filter ranged from $1$ to $256$ KB, with $32$ KB used for a $\uptau$ of $5$ ms. DSE (Section \ref{DSE}) can be performed to further optimize the memory size and configuration.



\begin{figure*}[htb]
\centering
\captionsetup[subfigure]{font=scriptsize,labelfont=scriptsize}
\subfloat[]{\includegraphics[width=0.33\textwidth, bb=0 0 2000 1400]{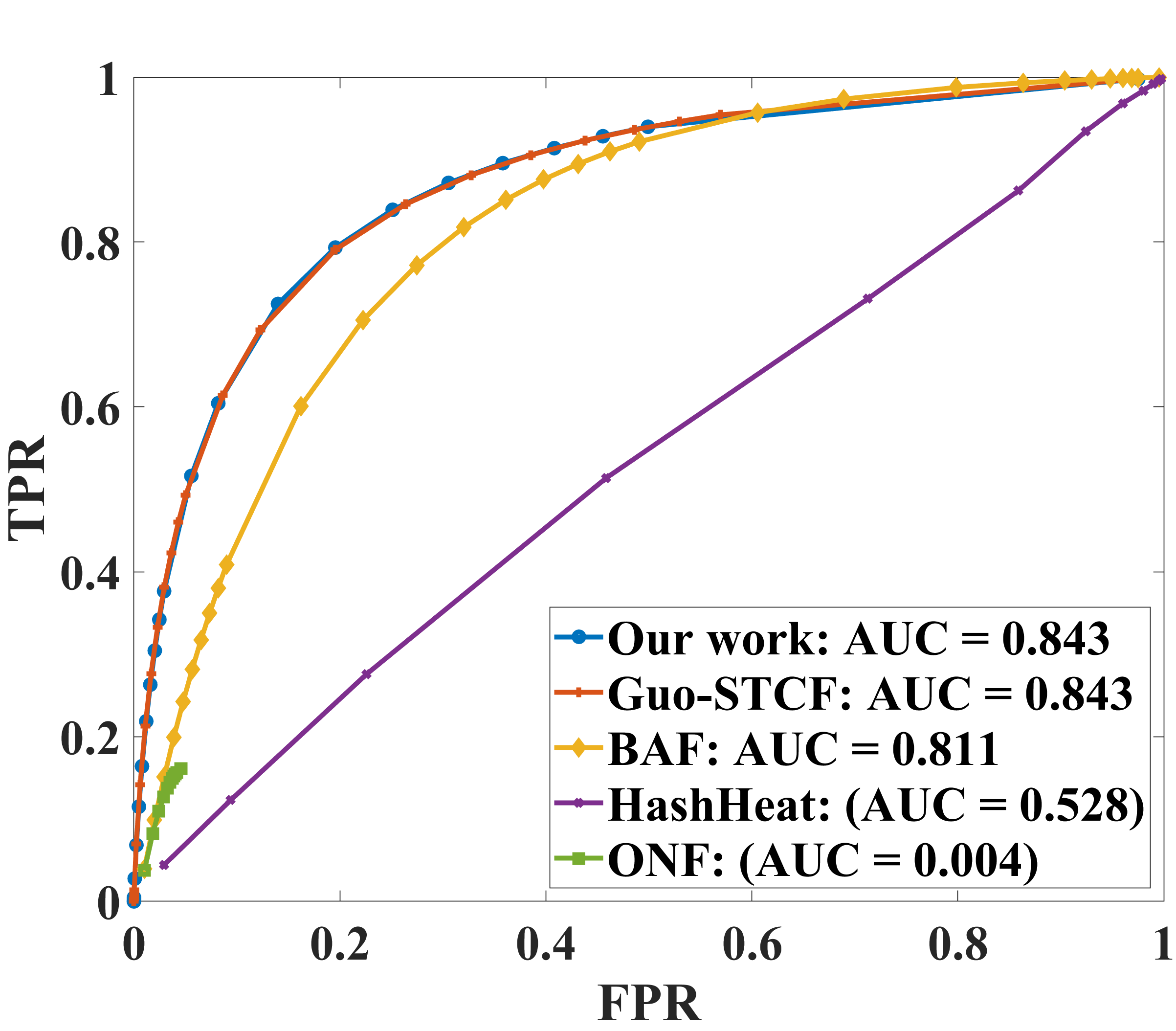}} 
\subfloat[]{\includegraphics[width=0.33\textwidth, bb=0 0 2000 1400]{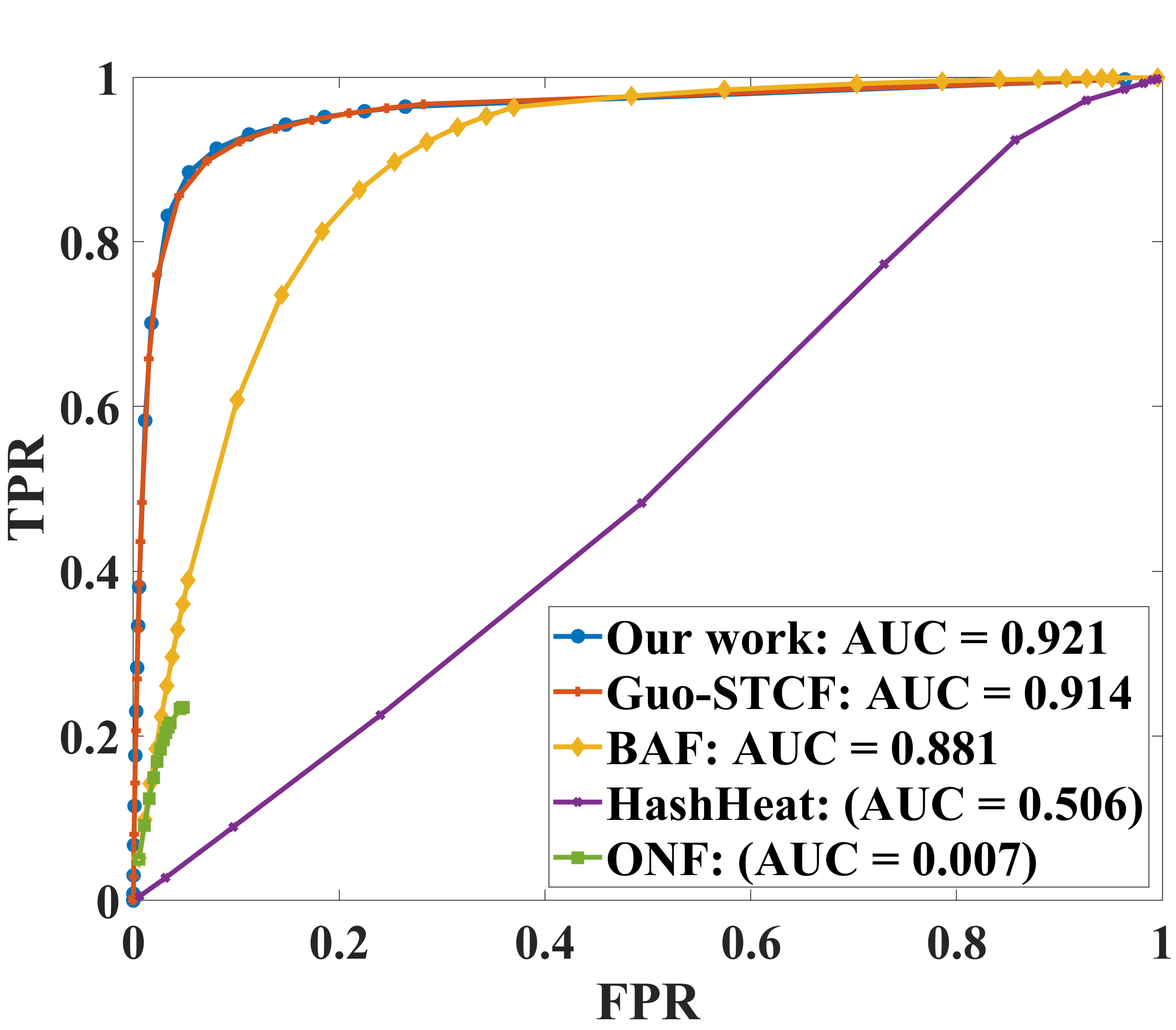}}
\subfloat[]{\includegraphics[width=0.33\textwidth, bb=0 0 2000 1400]{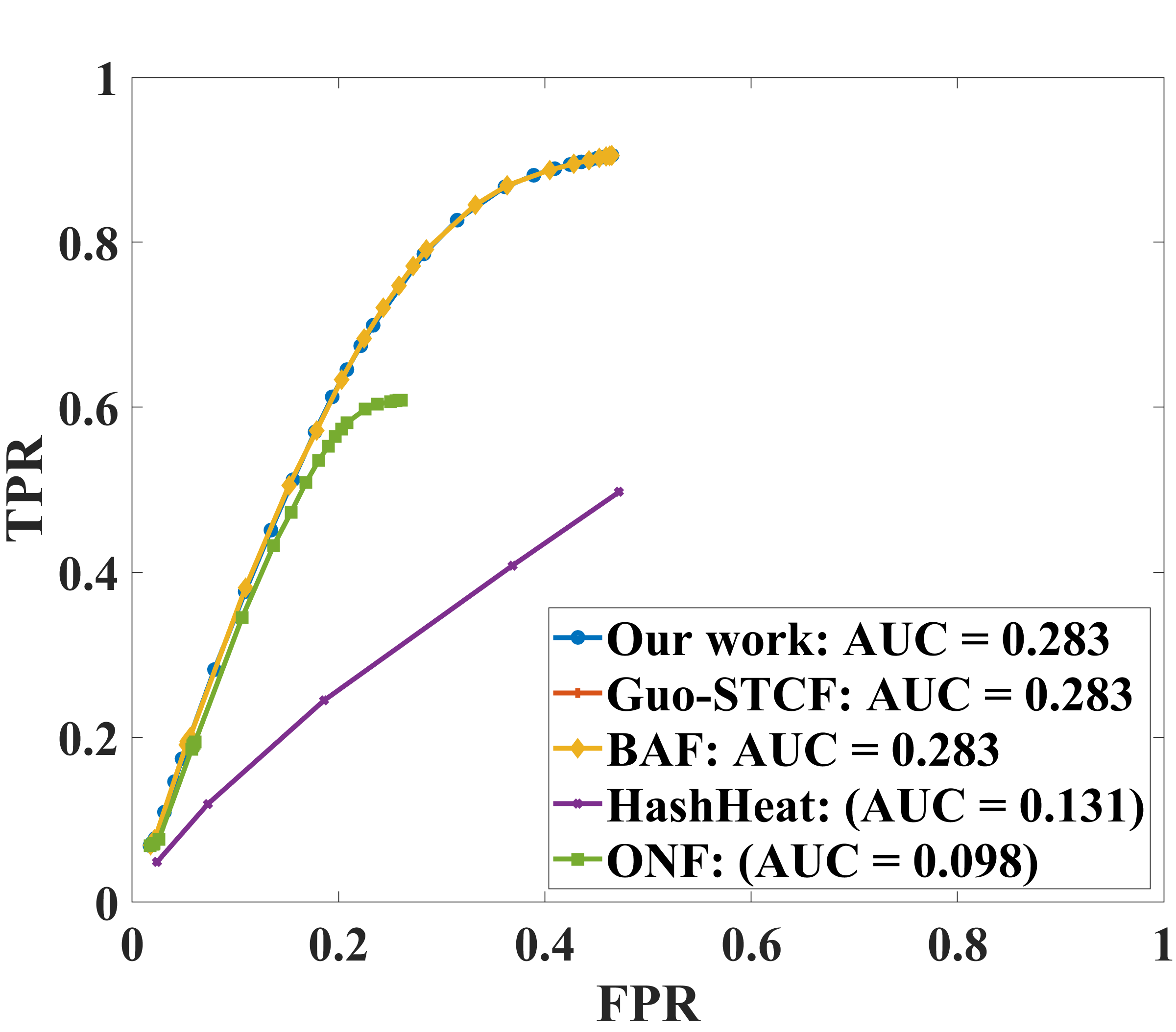}}
 \caption{FPR-TPR ROC curve and AUC of our filter  vs. other filters for (a) \textit{driving} data mixed with \textit{shot noise}, both from the DND21 dataset, (b) \textit{hotel-bar} data mixed with \textit{shot noise} from the DND21 dataset and (c) Traffic data containing naturally occurring sensor noise.}
\label{fig:fpr_vs_tpr}
\end{figure*}

Fig. \ref{fig:fpr_vs_tpr}(a) shows the ROC curves of the filters input with driving data mixed with shot noise (low intensity illumination condition), both from the DND21 dataset. For our filter and the Guo-STCF, the presence of four supporting events was set as the requirement for declaring an event as a signal. The results indicate that our filter and the Guo-STCF exhibit similar performances, exceeding the performances of the other filters. 

Fig. \ref{fig:fpr_vs_tpr}(b) shows the performances of the filters input with the \textit{hotel-bar} dataset from DND21 mixed with shot noise. The results are similar to those simulated using the driving dataset. The best performance was exhibited by the Guo-STCF and our filter.
\ifdefined \OnlineSM
\href{https://sites.google.com/view/bf2-event-based-filter/supplementary-material/section-e}{SM Section E}
\else
SM Section E
\fi
contains images that enable a qualitative comparison of the filtering efficacy of the different filters.

The results for filtering performance based on the \textit{traffic} dataset are shown in Fig. \ref{fig:fpr_vs_tpr}(c). Here again, our filter and the Guo-STCF are the top performers. Although the ONF exhibits a comparatively better performance with this dataset compared to the other datasets, it should be noted that the scenarios presented by this dataset for filtering are less demanding. In the ONF, one memory cell is shared by events from an entire row or column, with the possibility of it being overwritten by events unrelated to the current activity. Therefore, this filter is suited only for scenarios with sparse activity.

The above results show that our filter and the Guo-STCF display similar filtering performances, both performing better than the other filters that were considered. However, the hardware resource requirements of our filter are much lower than those of the Guo-STCF filter as demonstrated in Section \ref{hardware resource comparison}, making our filter better suited to near-sensor hardware implementations in resource-constrained IoT applications.



\subsection{Hardware Implementation Efficiency} \label{hardware resource comparison}

The proposed filter is designed to filter BA noise close to the sensor using limited hardware resources. Two factors that we consider in evaluating hardware implementation efficiency are \textit{energy per event} and \textit{memory requirement}, as they directly affect the power consumption and silicon area utilization of the circuits implemented in hardware. In the following sections, we evaluate and compare the filters based on these factors.

\subsubsection{Energy per event}
We estimated the energy consumed by the different filters to process a single incoming event. This was done by analyzing each filter's algorithm, listing all the basic operations needed to process an incoming event, estimating the energy required to execute each operation, and summing up the energy costs. Data from \cite{Horowitz2014} which provides the energy cost for different arithmetic and memory access operations in a $45$ nm, $0.9$ V power supply CMOS process was used to estimate the total energy required per event. The results are shown in Table \ref{tab:pJ_per_event}.
\begin{table} [htbp] 
\centering
\caption{Comparison of energy consumption (pJ per event) for different sensor sizes} \label{tab:pJ_per_event}
\begin{tabular}{|l|l|l|l|l|}
\hline
\textbf{\begin{tabular}[c]{@{}l@{}}\backslashbox{Filter}{Sensor\\Size}\end{tabular}} & \textbf{240$\times$180} & \textbf{346$\times$260} & \textbf{640$\times$480} & \textbf{1280$\times$960} \\ \hline
\textbf{Guo-STCF} & 140 & 206 & 514 & 1820 \\ \hline
\textbf{HashHeat} & {20} & {20} & {20} & {20} \\ \hline
\textbf{ONF} & 67 & 72 & 85 & 114 \\ \hline
\textbf{Our work} & 29 & 29 & 31 & 34 \\ \hline
\end{tabular}
\end{table}
The detailed calculations used to arrive at the values shown in Table \ref{tab:pJ_per_event} are given in
\ifdefined \OnlineSM
\href{https://sites.google.com/view/bf2-event-based-filter/supplementary-material/section-c}{SM Section C}.
\else
SM Section C.
\fi
\subsubsection{Memory requirement} 
\begin{figure*}[htb]
\centering 
\captionsetup[subfigure]{font=scriptsize,labelfont=scriptsize}
\subfloat[] {\includegraphics[width=0.33\textwidth, bb=0 0 1100 500]{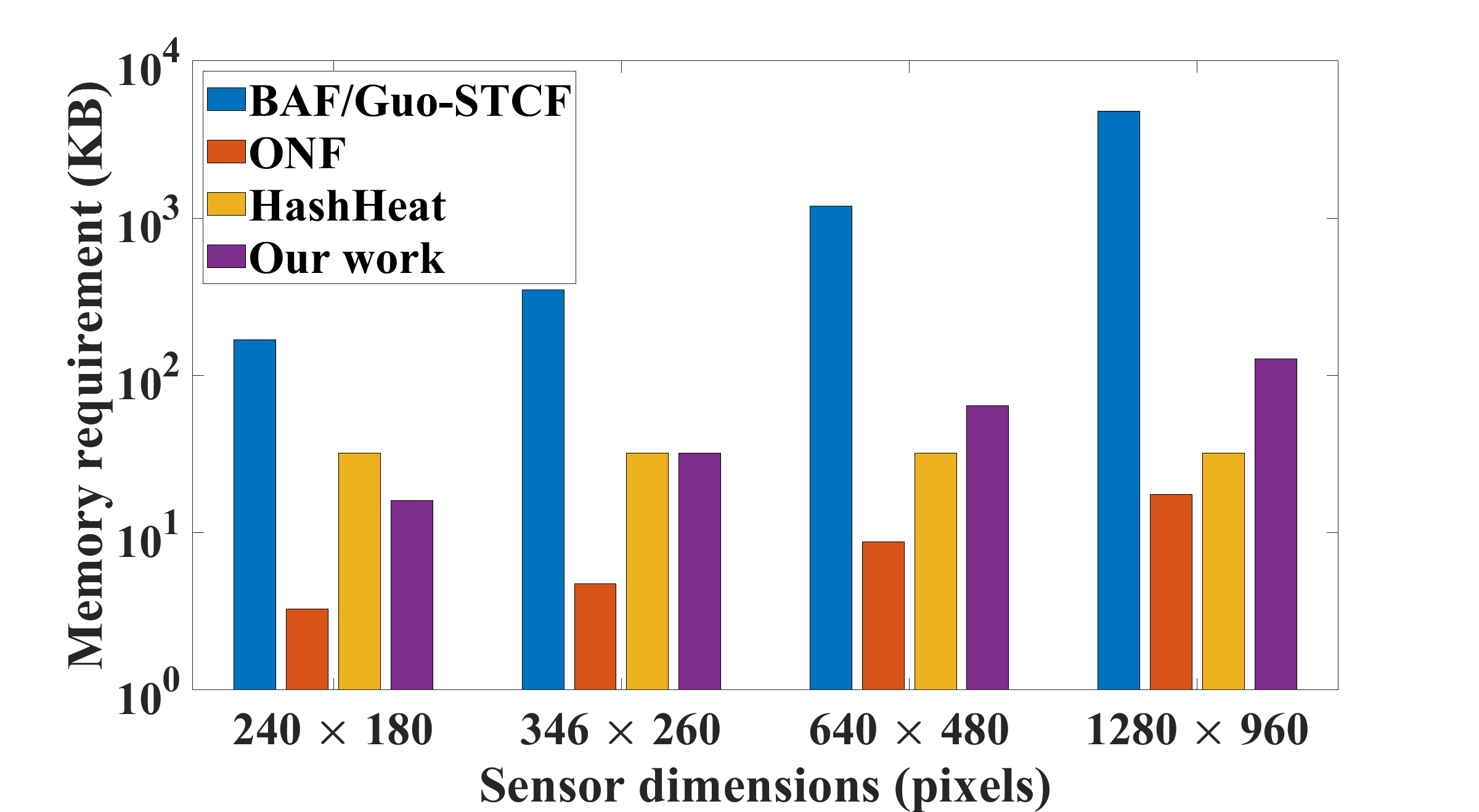}} 
\subfloat[] {\includegraphics[width=0.33\textwidth, bb=0 0 1100 500]{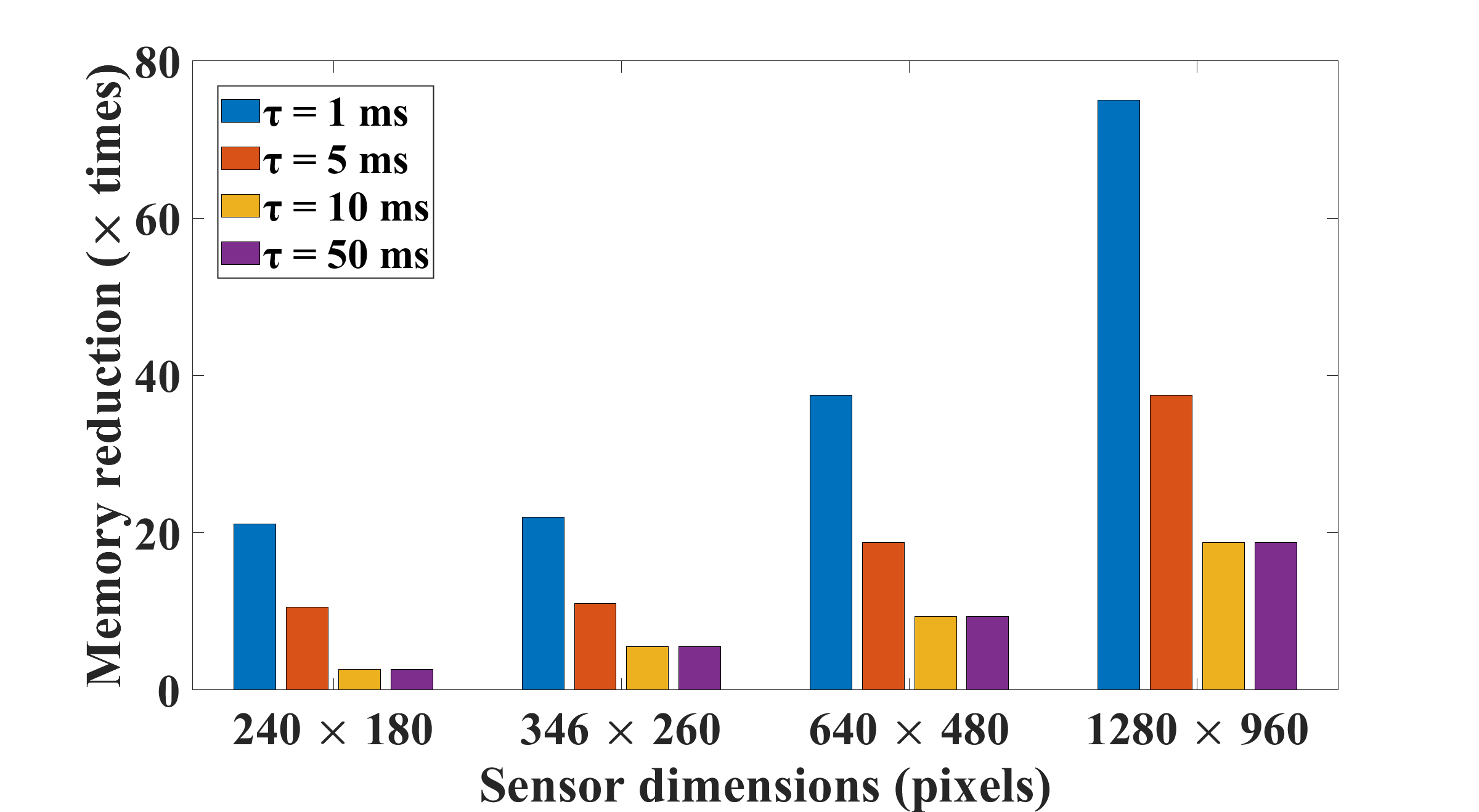}} 
\subfloat[] {\includegraphics[width=0.33\textwidth, bb=0 0 1100 500]{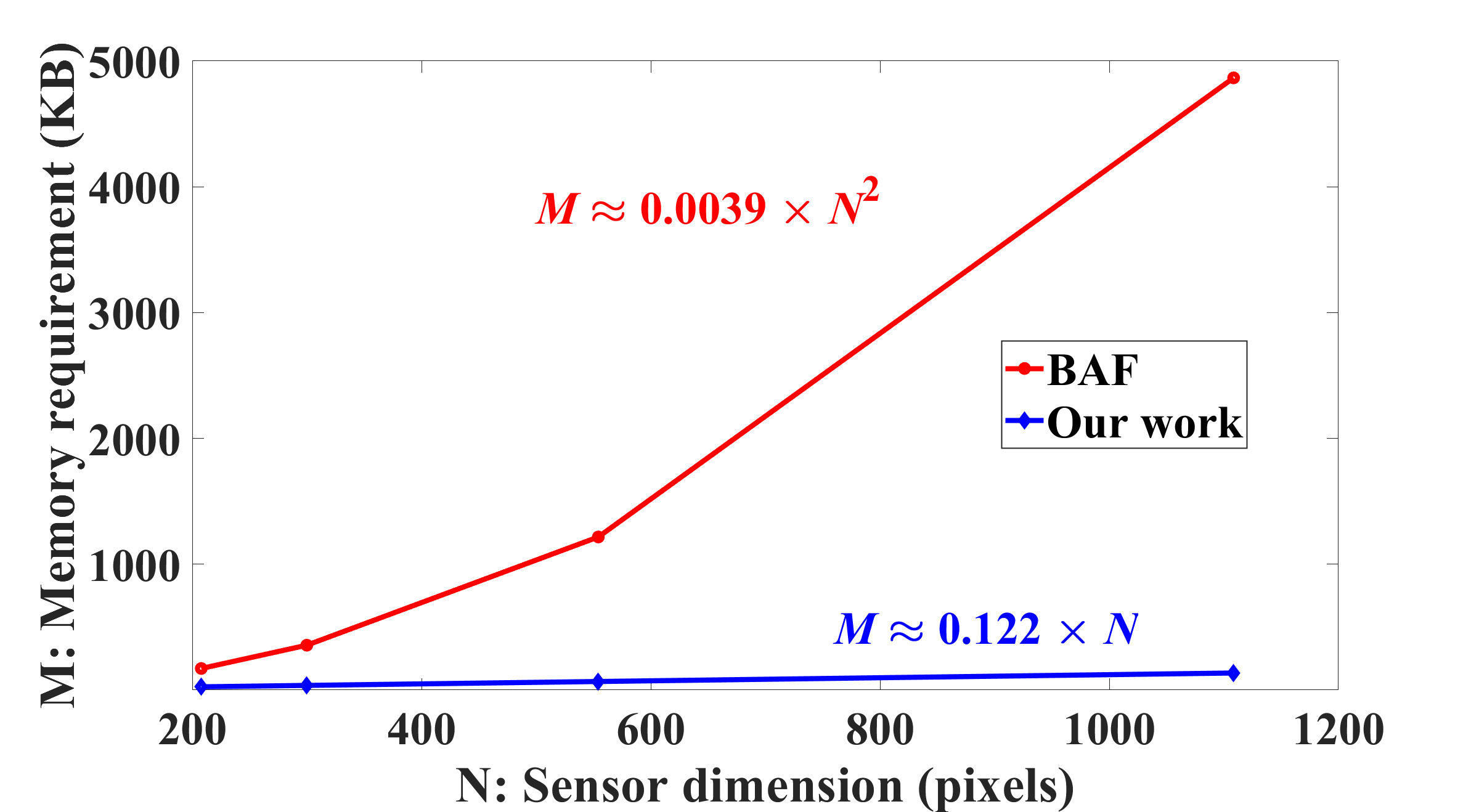}} 
\caption{(a) Memory requirement of compared filters for different sensor sizes - memory requirement of our filter is much less than that of the Guo-STCF/BAF across all sensor resolutions ($\tau = 5$ ms). 
(b) Comparison of memory reduction achieved with our filter with those of BAF/Guo-STCF for different sensor sizes and $\uptau$ values.
(c) Scaling of memory requirement with sensor size for our filter compared to BAF/Guo-STCF - scaling in our proposed filter is linear while it is quadratic in the BAF/Guo-STCF.
}
\label{fig:memory_savings} 
\end{figure*}

Fig. \ref{fig:memory_savings}(a) shows the estimated memory requirement for the different filters, for different sensor sizes and a $\uptau$ of $5$ ms.  
For the Guo-STCF, BAF, and ONF, the memory requirements can be directly calculated from the sensor dimensions as explained in Section \ref{BAF} and Section \ref{ONF}. For our filter and HashHeat, the optimum memory usage was determined through theoretical analysis (see Section \ref{Theoretical Analysis of filter}), and simulations using versions of the \textit{driving} dataset mixed with shot noise, scaled appropriately for sensor sizes other than $346$ $\times$ $260$. It can be observed that the memory requirement for our filter is closer to that of the ONF or Hashheat and much less than that of the BAF/Guo-STCF. 

Fig. \ref{fig:memory_savings}(b) shows the typical reduction in memory requirements achieved using our filter as compared to the BAF/Guo-STCF, for varying values of $\uptau$; the details of the calculation are given in
\ifdefined \OnlineSM
\href{https://sites.google.com/view/bf2-event-based-filter/supplementary-material/section-d}{SM Section D}.
\else
SM Section D.
\fi
It can be seen that the advantage of our filter with respect to memory increases significantly with increasing sensor dimensions. This is desirable as larger sensor dimensions is the trend with the latest developments in image sensor design \cite{Gallego2019}. The advantage of reduced memory nevertheless decreases with larger $\uptau$ requirements; however, $\uptau>10$ ms rarely needs to be used for noise filtering, and our proposed method can be expected to provide significant benefits within the commonly used parameter regime.

Fig. \ref{fig:memory_savings}(c) shows the increase in memory requirements of the BAF and our filter with increasing sensor sizes, for a $\uptau$ of $5$ ms. In this figure, $N$ is the geometric mean of the two sensor dimensions and $M$, the memory requirement in KB. The equations shown in the graph approximate the memory requirements of the BAF and our filter for different values of $N$. It may be observed that the BAF (and by extension, the Guo-STCF) has a memory requirement that is quadratically related to the sensor size while with our filter, this relationship is linear. For the driving dataset used and the range of $\uptau$ values simulated ($1$--$50$ ms), we found that the memory requirement of our filter can be approximated by the relation $M = F \times N$ where $F$ ranges from $0.06$ to $0.22$.

\begin{figure} [htbp]
\centering 
\includegraphics[width=0.75 \columnwidth, bb=0 0 1200 750]{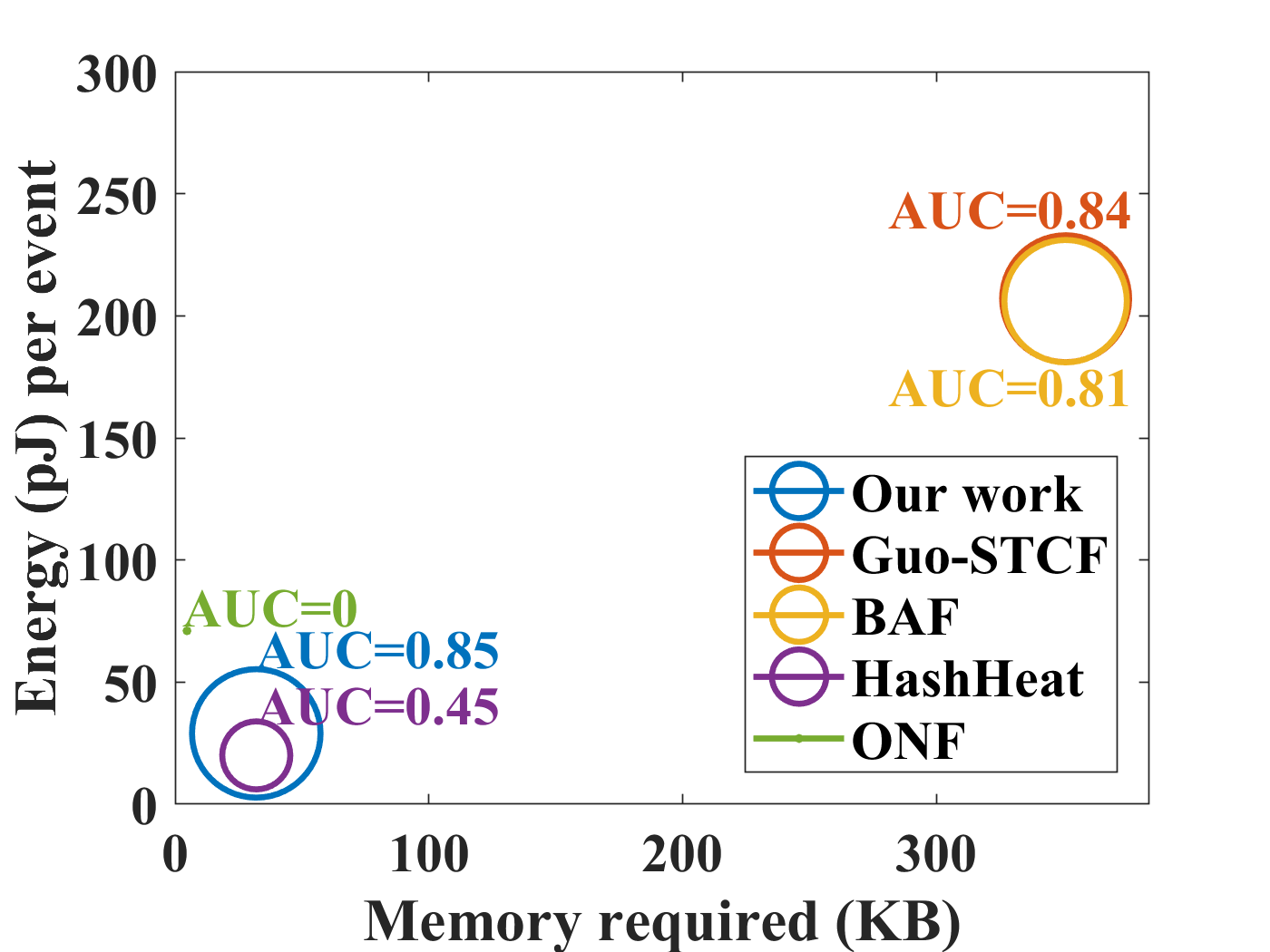}
\caption{Memory requirement, energy dissipation per event and AUC, derived from simulations using \textit{driving} dataset mixed with \textit{shot} noise; $\uptau$ = $5$ ms, sensor size = $346$ $\times$ $260$ pixels.}
\label{fig:combined_auc_pj_mem}
\end{figure}

The combined effect of filter performance, energy dissipation per event, and memory requirement of the filters is illustrated in Fig. \ref{fig:combined_auc_pj_mem} which is based on simulations using the \textit{driving} dataset mixed with \textit{shot} noise, target $\uptau$ of $5$ ms and a sensor size of  $346$ $\times$ $260$ pixels. The circles represent the different filters, their areas being proportional to the corresponding AUC values. The $x$ and $y$ coordinates of the centers of the circles represent the corresponding energy (pJ) dissipation per event and memory requirement (KB) respectively. Circles close to the origin represent filters with higher hardware efficiency (those with lesser energy and memory requirements) while bigger circles indicate better filtering performance. For the chosen sensor size, the memory requirement for our filter is $32$ KB. From this figure, it is clear that the proposed filter  provides high accuracy similar to the BAF/Guo-STCF, but requires several orders less of memory and energy---similar to the ONF or HashHeat. Although the ONF requires less memory than the proposed filter, its performance is the lowest among all the filters considered. Similarly, while HashHeat has a lower memory requirement and a higher throughput than that of our filter, poor performance rules out its use under complex, real-life scenarios.
\section{Hardware Implementation and Discussion} \label{Hardware}

\subsection{Hardware Features}
The proposed filter using the $BF_2$ was implemented in hardware to further assess potential advantages over other filters. The following features of the proposed filter's architecture/design makes it suitable for efficient hardware implementation:

\begin {enumerate}
\item{Binary logic operations:
Most of the computations performed by the filtering algorithm are simple binary logic operations that are efficiently implemented in hardware. 
As a result, the computational energy per event and the hardware resources required are minimized. The only arithmetic operation performed in the filter is a unary increment by the row counter. Since the data-paths are all single-bit, there is no loss of precision when the algorithm is mapped to hardware. There is agreement between the results of MATLAB simulation and post-implementation Vivado simulation.}

\item{Hash-based searching:
The STCF search operation mentioned in Section \ref{BA Noise} is performed in the filter using the $H_3$ hash function \cite{Ramakrishna1994} using binary logic and no arithmetic operations. This makes the hashing logic fast and compact when implemented in hardware, as shown in 
\ifdefined \OnlineSM
\href{https://sites.google.com/view/bf2-event-based-filter/supplementary-material/section-a}{SM Section A}.
\else
SM Section A.
\fi
This is particularly advantageous as the hash look-up needs to be performed for every incoming event.}

\item{Elimination of $32$-bit timestamps for classification: As described in Section \ref{Proposed Filter}, events are stored in the rows of the $BF_2$ array in the filter, based on their arrival time; the $32$-bit time stamp in the AER event packet is not directly stored or used by the filter, which greatly reduces the overall memory requirement. Also, $32$-bit arithmetic operations involving the time stamps are avoided, thus substantially saving computational energy.}


\end {enumerate}

\subsection{Hardware Implementation Considerations}

\subsubsection{Memory organization} \label{memory_org}

In hardware implementations, the memory within the filter is organized based on functional requirements, the objective being to maximize filter throughput and minimize energy consumption.

\begin{figure} [htb] \centering \includegraphics[width= \columnwidth, bb= -100 50 1000 280] {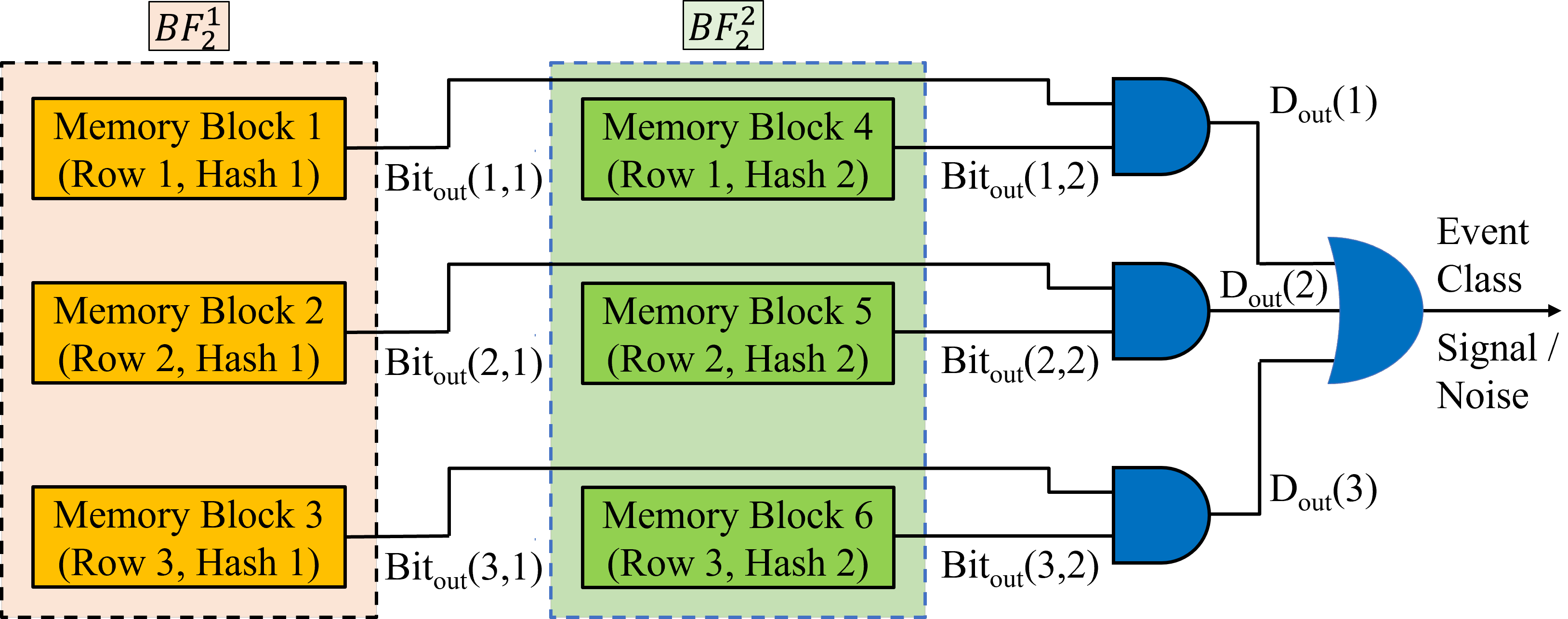} 
\caption{Memory organization for FPGA-based realization of proposed filter.}
\label{fig:memory_organization}
\end{figure}

The entire memory is organized as a $D \times K$ array of memory blocks where $D$ is the depth of array and $K$ is the number of hash functions used. Each column of this array stores the incoming events, as explained in Section \ref{Proposed Filter}. Each row of the array is a set of $K$ memory blocks, each of which is internally organized as a $1 \times W$ bit array. Significantly, each memory block can be accessed independently so that it is possible to read from and write to all memory blocks in parallel.
Fig. \ref{fig:memory_organization} shows an example of memory organization in a filter with $3$ rows and $2$ hash functions ($D$ = $3$, $K$ = $2$).

This memory structure makes it possible to store an incoming event by setting bits in all memory blocks of the currently active row of all FIFOs concurrently. It is also possible to search for support from one of the neighboring cells of an incoming event, by reading all memory blocks in the entire array in one cycle. To search for the presence of support from all $8$ neighbors of an event in the array, $8$ cycles are needed. The AND gates in Fig. \ref{fig:memory_organization} search for a supporting event in one row as a whole, and the OR gate combines the results from all the rows to decide if the event is signal or noise.

\subsubsection{Memory row reset} \label {mem_reset}

As mentioned in Section \ref{Data_Structure}, a portion of the $BF_2$ array has to be cleared periodically to prevent the filter from saturating and causing a large number of false positives. This is addressed by Algorithm \ref{alg:classification} in a single step. However, it is not possible to implement this in hardware as the entire memory block cannot be cleared in a single step. Therefore, in practice, we reset a row of the FIFOs over several cycles. While the `active row' in the $BF_2$ array is being written into, the next row in sequence is simultaneously cleared.


\textbf{}It is necessary to make sure that the clearing of a row is completed in the time allotted per row, ($\uptau_{row}$) even when the bit array width $W$ is large. To speed up the memory reset process, we use memory organized as multi-bit words that can be written with zeros to clear all the bits in a word simultaneously. By adding peripheral logic to this multi-bit word memory, it is also possible to set and read single bits during event storage and search, as required by the filter data structure.


In summary, the memory, and the logic around it is designed such that:\\
1. Multiple bits can be cleared simultaneously during the bit clearing phase.
2. Individual bits can be set and read during the write and read phases respectively.

\subsection{FPGA Implementation and Comparisons}
To estimate the hardware resource requirements of the proposed filter and compare it with other filters from \cite{Guo2020} \cite{Khodamoradi2021}, we implemented the filter design on an FPGA device. We selected the XC7A35T FPGA device from AMD-Xilinx to target our design. This device belongs to the Artix-7 family of devices that includes compact devices with high performance-per-watt ratios, making them suitable for use in sensor interface applications where low-power and small footprints are desirable. All filters were implemented on the same family of devices.

Table \ref{tab:fpga_resources} shows the resource utilization of the selected device with the filter core implemented on it. The filter parameters were $W = 16384, D = 4, K = 4$ and $\uptau = 5$ ms.
For the selected FPGA device, the maximum clock frequency achieved was $\approx$ $166$ MHz. As our filter takes 8 cycles to perform a search in the spatio-temporal neighborhood to classify an event, and an additional cycle to store it in the array, the throughput achieved is ($1/9$)th of the clock frequency or $\approx$ $18$ Mevents-per-second (Meps) with a latency of $\approx$ $54$ ns.

\begin{table}[htbp]
\centering
\caption{FPGA resource utilization of the proposed filter}
\label{tab:fpga_resources}
\begin{tabular}{|l|r|r|r|l}
\cline{1-4}
\textbf{FPGA Resource} & \multicolumn{1}{l|}{\textbf{Used}} & \multicolumn{1}{l|}{\textbf{Available}} & \multicolumn{1}{l|}{\textbf{Utilization}} &  \\ \cline{1-4}
BRAM & 8 & 50 & 16\% &  \\ \cline{1-4}
FF & 71 & 41600 & 0.17\% &  \\ \cline{1-4}
LUT & 1067 & 20800 & 5.13\% &  \\ \cline{1-4}
DSP & 0 & 90 & 0\% &  \\ \cline{1-4}
\end{tabular}
\end{table}

Table \ref{tab:fpga_resources_all} compares the FPGA resource utilization, power and throughput of the different filters when implemented on an XC7A35T FPGA device. The FPGA resource utilization and throughput figures for the BAF and ONF were derived from \cite{Khodamoradi2021} where their power consumption values are not reported. The FPGA resource utilization, power and throughput values of HashHeat were taken from \cite{Guo2021}. To estimate the power consumption of our filter, we generated a Switching Activity Interchange Format (SAIF) file from a post-layout timing simulation of the filter. This file was then fed into the Vivado power estimator tool to estimate its power consumption with a high confidence level.

Further details related to the FPGA implementation are given in
\ifdefined \OnlineSM
\href{https://sites.google.com/view/bf2-event-based-filter/supplementary-material/section-f}{SM Section F}.
\else
SM Section F.
\fi
%

\begin{table}[htbp]
\centering
\caption{FPGA resource utilization, power and throughput comparison}
\label{tab:fpga_resources_all}
\begin{tabular}{|c|r|r|r|r|r|r|}
\hline 
\textbf{Filter} &
  \multicolumn{1}{c|}{\textbf{\begin{tabular}[c]{@{}c@{}}BRAM \\ (\%)\end{tabular}}} &
  \multicolumn{1}{c|}{\textbf{\begin{tabular}[c]{@{}c@{}}FF\\ (\%)\end{tabular}}} &
  \multicolumn{1}{c|}{\textbf{\begin{tabular}[c]{@{}c@{}}LUT\\ (\%)\end{tabular}}} &
  \multicolumn{1}{c|}{\textbf{\begin{tabular}[c]{@{}c@{}}DSP\\ (\%)\end{tabular}}} &
  \multicolumn{1}{c|}{\textbf{\begin{tabular}[c]{@{}c@{}}Power\\ (W)\end{tabular}}} &
  \multicolumn{1}{c|}{\textbf{\begin{tabular}[c]{@{}c@{}}Through-\\ put \\ (Meps)\end{tabular}}} \\ \hline
\textbf{BAF/STCF} & 363.39 & 0.40 & 2.27  & 0.00  & N.A  & 14.00 \\ \hline
\textbf{ONF}      & 4.00   & 2.07 & 4.66  & 0.00  & N.A  & 3.00  \\ \hline
\textbf{HashHeat} & 1.00   & 6.11 & 50.00 & 22.22 & 0.47 & 100   \\ \hline
\textbf{Our work} & {16.00}  & 0.17 & 5.13  & 0.00  & 0.22 & 18.00 \\ \hline
\end{tabular}
\end{table}

The throughput of our filter is high enough to process output from DVS cameras such as the DVS$128$ \cite{Lichtsteiner2008}, DAVIS$240$ \cite{Brandli2014} and DAVIS$346$ \cite{Gallego2019} in real-time. However, the current filter design cannot be directly used to filter output of cameras with peak throughput higher than $18$ Meps.

To improve the filter's throughput, it is possible to augment the current architecture with a combination of FIFOs and multi-ported memories.

FIFOs can be used to buffer the peak output from the camera to prevent data loss. While currently, some DVS cameras have high peak output event rates, the average event rate in practical situations tend to be much lower. For example, in the driving dataset, the peak event rate including noise is 272 Meps but the average rate is only 1.11 Meps. In such scenarios, a FIFO between the camera and our filter will be able to handle higher throughput, although at the expense of the added FIFO memory and the resulting increase in latency.

As mentioned in Section \ref{memory_org}, the processing time required by our filter for event classification is mostly to search for supporting events at the eight locations neighbouring the current event. In custom design, the creation of multi-port memories can be considered, to enable parallel search for multiple events per cycle, thereby reducing the processing time and increasing the throughput.


\section{Conclusion and Future Work}
In this paper, we introduce a new data structure termed $BF_2$ for storing sparse events generated from dynamic vision sensors.
Many event-based algorithms such as noise filtering and corner detection involve searching for the presence of past events within a specified spatio-temporal volume; this can easily be accomplished using this data structure. Compared to other commonly used data structures such as the TS which require memories of the order of sensor resolution, our proposed method exploits the sparsity of event data and uses hashing to store the data in a compressed 2-D array. To overcome saturation problems in the conventional BF, or data loss during reset of HashHeat, $BF_2$ deletes a single row after every $\tau_{row}$ time period and starts storing data there.

To demonstrate the efficacy of the proposed data structure, we used it to implement a BA noise filter.  Experimental results show that the proposed filter is able to perform noise filtering with high efficacy in complex scenarios involving moving cameras and challenging lighting conditions where other hardware-friendly filters like the ONF and HashHeat fail. Compared to other high-performance filters like the BAF/Guo-STCF, our proposed filter requires $\approx$ $11$--$38$ $\times$ less memory and $\approx$ $7$--$53$ $\times$ less energy per event across different sensor sizes for $\tau=5$ ms. These features namely low-energy and low-memory-area make it suitable for implementation in always-On edge intelligence devices to enable interfacing with DVS.

In future, we plan to implement a hardware-friendly, event-based corner detection scheme based on the proposed data structure, followed by the development of an event-based tracker optimized for resource-constrained embedded systems. We also plan to introduce further improvements in the digital implementation of the filter to improve its throughput.



\section*{Acknowledgements}

This research is part of the programme DesCartes and P. K. Gopalakrishnan and C.H. Chang are supported by the National Research Foundation, Prime Minister’s Office, Singapore under its Campus for Research Excellence and Technological Enterprise (CREATE) programme. 

A. Basu is supported by Grant No. 9380132 from City University of Hong Kong.

\section*{Supplementary Material}\label{Supplementary}
Further information related to this paper (such as filter output qualitative comparisons, resource calculations and FPGA implementation results) is provided in the
\ifdefined \OnlineSM
Supplementary Material available at the following link: \url{https://sites.google.com/view/bf2-event-based-filter/supplementary-material}.
\else
attached Supplementary Material.
\fi
\bibliographystyle{IEEEtran}
\bibliography{Bibliography.bib}
\newpage
\vskip -2\baselineskip plus -1fil 
\begin{IEEEbiography}
[{\includegraphics[width=1in,height=1.25in,clip,keepaspectratio]{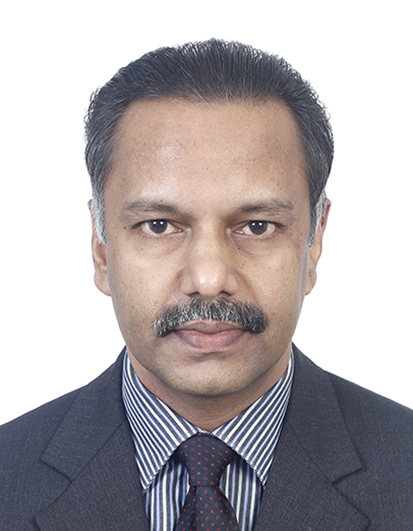}}]
{Pradeep Kumar Gopalakrishnan\ }(M'01-SM’10) received his B.Tech degree in Electrical Engineering from the University of Kerala, India in 1989 and his M.Tech degree in Electronics Design and Technology from the Indian Institute of Science in 1997. He has over 25 years of experience in the industry, mainly in ASIC and embedded systems design. He worked in organizations such as Philips Semiconductors, Broadcom, Institute of Microelectronics (A*STAR), Siemens Corporate Research and Xilinx before joining NTU. He is currently pursuing a PhD degree at Nanyang Technological University, Singapore. His research interests include low-power Machine Learning architectures and Neuromorphic Systems.
\end{IEEEbiography}
\vskip -2\baselineskip plus -1fil
\begin{IEEEbiography}
[{\includegraphics[width=1in,height=1.25in,clip,keepaspectratio]{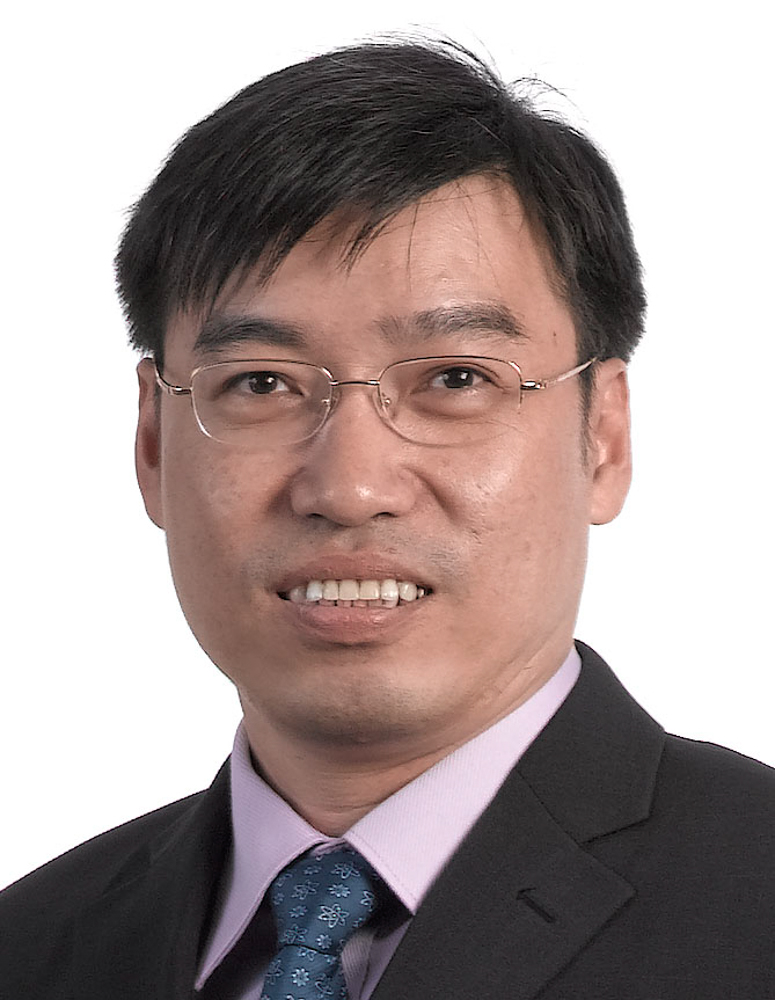}}]
{Chang Chip Hong\ } (M’98-SM’03-F'18) received the B.Eng. (Hons.) degree from the National University of Singapore in 1989, and the M. Eng. and Ph.D. degrees from Nanyang Technological University (NTU) of Singapore in 1993 and 1998, respectively. He is a Professor of the School of Electrical and Electronic Engineering (EEE) of NTU. He held joint appointments with the university as Assistant Chair of Alumni from 2008 to 2014, Deputy Director of the Center for High Performance Embedded Systems from 2000 to 2011, and Program Director of the Center for Integrated Circuits and Systems from 2003 to 2009. He was conferred the 2022 VISTA award of excellence in hardware security. He has co-edited 6 books, published 13 book chapters, more than 100 international journal papers (more than 80 are in IEEE) and more than 180 refereed international conference papers (mostly in IEEE), and delivered more than 50 keynotes, tutorial and invited seminars. His current research interests include hardware security, AI security, trustworthy sensing and hardware accelerators for post-quantum cryptography and edge computational intelligence.
            
Dr. Chang currently serves as the Senior Area Editor of IEEE Transactions on Information Forensic and Security, and Associate Editor of the IEEE Transactions on Circuits and Systems-I and IEEE Transactions on Very Large Scale Integration (VLSI) Systems. He also served as the Associate Editor of the IEEE Transactions on Information Forensic and Security and IEEE Transactions on Computer-Aided Design of Integrated Circuits and Systems from 2016 to 2019, IEEE Access from 2013 to 2019, IEEE Transactions on Circuits and Systems-I from 2010 to 2013, Integration, the VLSI Journal  from 2013 to 2015, Springer Journal of Hardware and System Security from 2016 to 2020 and Microelectronics Journal from 2014 to 2020. He guest edited around 10 special issues and served in the organizing and technical program committee of more than 70 international conferences (mostly IEEE). He is an IEEE Fellow, IET Fellow, and 2018-2019 Distinguished Lecturer of IEEE Circuits and Systems Society.

\end{IEEEbiography}
\vskip -2\baselineskip plus -1fil
\begin{IEEEbiography}
[{\includegraphics[width=1in,height=1.25in,clip,keepaspectratio]{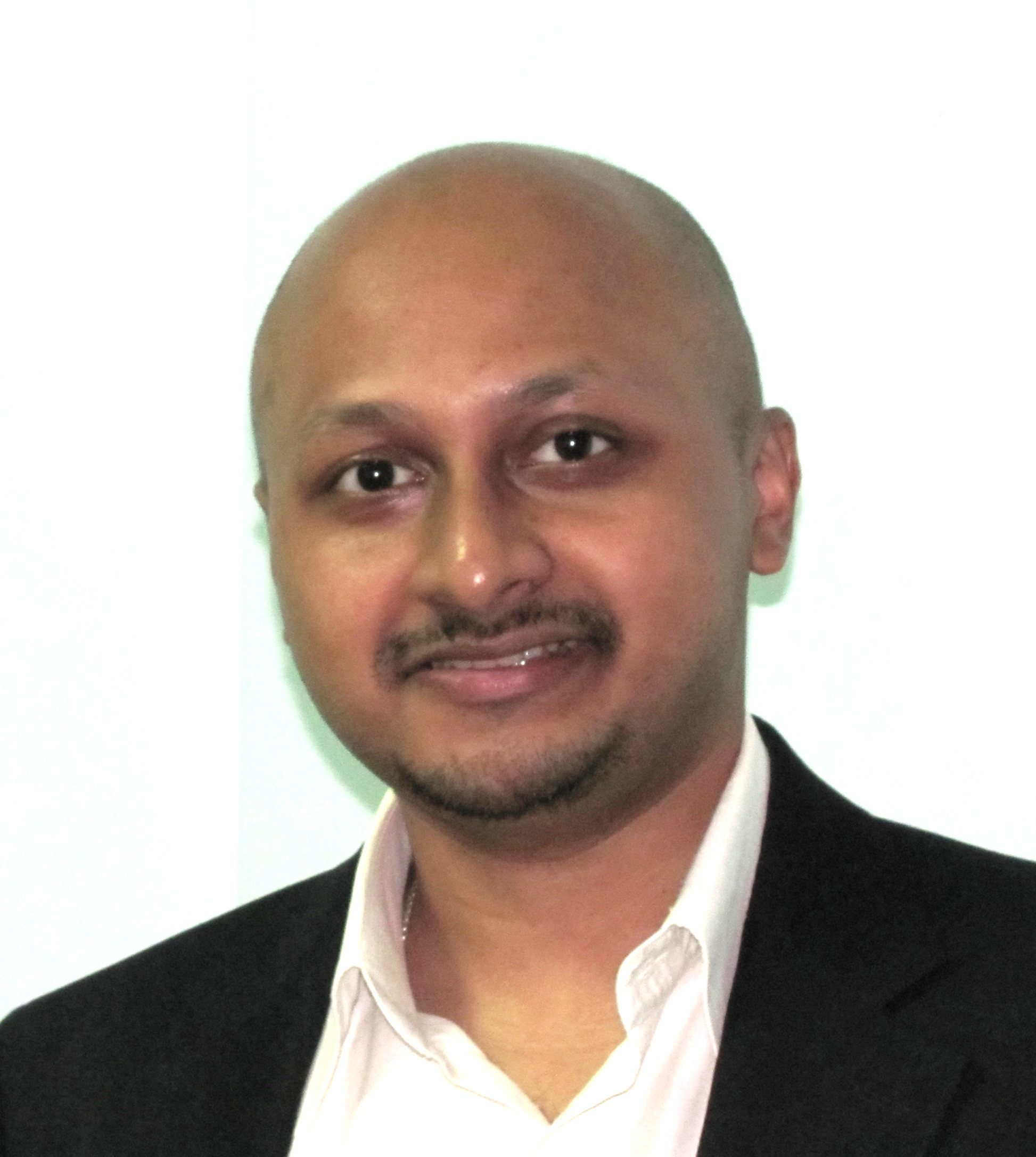}}]
{Arindam Basu\ }(M10-SM’17) received his B.Tech and M.Tech degrees in Electronics and Electrical Communication Engineering from the Indian Institute of Technology, Kharagpur in 2005, the M.S. degree in Mathematics and PhD. degree in Electrical Engineering from the Georgia Institute of Technology, Atlanta in 2009 and 2010 respectively. Dr. Basu received the Prime Minister of India Gold Medal in 2005 from I.I.T Kharagpur. He is currently a Professor in City University of Hong Kong in the Department of Electrical Engineering and was a tenured Associate Professor at Nanyang Technological University before this. He is currently an Associate Editor of IEEE Sensors journal, Frontiers in Neuroscience and IEEE Transactions on Biomedical Circuits and Systems. He has served as IEEE CAS Distinguished Lecturer for 2016-17 period.

Dr. Basu received the best student paper award at Ultrasonics symposium, 2006, best live demonstration at ISCAS 2010 and a finalist position in the best student paper contest at ISCAS 2008. He was awarded MIT Technology Review's inaugural TR35@Singapore award in 2012 for being among the top 12 innovators under the age of 35 in SE Asia, Australia and New Zealand. He is a technical committee member of the IEEE CAS societies of Biomedical Circuits and Systems, Neural Systems and Applications (Chair) and Sensory Systems. His research interests include bio-inspired neuromorphic circuits, non-linear dynamics in neural systems, low power analog IC design and programmable circuits and devices.
\end{IEEEbiography}

 




\end{document}